\documentclass[11pt,eqs]{article}
\usepackage{latexsym}

\def\cleq{\setcounter{equation}{0}}
\begin{document}
\title{Noncommutativity in weakly curved background by canonical methods
\thanks{Work supported in part by the Serbian Ministry of Science and
Technological Development, under contract No. 171031.}}
\author{Lj. Davidovi\'c \thanks{e-mail address: ljubica@ipb.ac.rs} and B. Sazdovi\'c
\thanks{e-mail address: sazdovic@ipb.ac.rs}\\
{\it Institute of Physics,}\\
{\it University of Belgrade,}\\
{\it 11001 Belgrade, P.O.Box 57, Serbia}}
\maketitle
\begin{abstract}
Using the canonical method,
we investigate the $Dp-$brane world-volume noncommutativity
in a weakly curved background.
The term ``weakly curved'' means that,
in the leading order,
the source of non-flatness is infinitesimally small Kalb-Ramond field $B_{\mu\nu}$,
linear in coordinate,
while the Ricci tensor does not contribute,
being the infinitesimal of the second order.
On the solution of boundary conditions,
we find a simple expression for the space-time coordinates
in terms of the effective coordinates and momenta.
This basic relation helped us to
prove that noncommutativity appears only on the world-sheet boundary.
The noncommutativity parameter
has a standard form,
but with infinitesimally small and coordinate dependent antisymmetric tensor $B_{\mu\nu}$.
This result coincides with that obtained on the group manifolds
in the limit of the large level $n$ of the current algebra.
After quantization, the algebra of functions on Dp-brane world-volume
is represented with the Kontsevich star product instead of the Moyal one
in the flat background.
\end{abstract}
\section{Introduction}
Quantization of the open string ending on Dp-branes has been
studied in many papers \cite{CDS}-\cite{ARS}. In the presence of
Kalb-Ramond antisymmetric tensor field $B_{\mu\nu}$, the Dp-brane
becomes noncommutative manifold.

In the simplest case, all background fields:
the metric tensor $G_{\mu\nu}$, the antisymmetric tensor
$B_{\mu\nu}$ and the dilaton field $\Phi$ are constant.
Geometrically, it corresponds to an embedding a flat Dp-brane into a flat background.
In that case, dilaton field does not give any contribution,
and the quadratic action represents two-dimensional free field theory.
The constant $B_{\mu\nu}$ field does
not affect dynamics in the world sheet interior. It contributes only to its boundary
and it is a source of noncommutativity.
Several methods have been used to investigate this case:
the operator product expansion of the open string vertex
operator \cite{Sch,VS},
the mode expansion of the classical solution \cite{AC},
the methods of conformal field theory
\cite{SW} and
the canonical quantization for constrained systems \cite{ACL,SN}.

In Refs. \cite{BKM,SN}, the inclusion of a dilaton field,
linear in space-time coordinates, has been investigated.
Because only the gradient of the dilaton field appears in space-time field equations,
this case technically behaves similarly to that with a constant background.
The dilaton field induces a commutative Dp-brane coordinate
in the direction of the dilaton gradient $\partial_\mu \Phi$.
For some particular relation between background fields, when $\partial_\mu \Phi$ is lightlike vector with respect
to the open- or closed-string metrics,
the local gauge symmetries appear.
They turn some Neumann boundary conditions into Dirichlet ones
and decrease the number of Dp-brane dimensions \cite{SN}.

In Refs. \cite{JdB},
the noncommutative properties of Dp-brane world volume embedded
in space-time of IIB superstring theory
has been investigated.
Similarly as in the bosonic theory,
the presence of $\sigma-$antisymmetric fields leads
to noncommutativity of the supercoordinates.
In the case of IIB theory, this supermultiplet beside $B_{\mu\nu}$
from NS-NS (Naveu-Schwarz) sector contains the difference of two gravitons,
$\psi^{\alpha}_{-\mu}$
from the NS-R (Ramond) sector
and the symmetric part of bispinor $F^{\alpha\beta}$ from the R-R sector.

In all previous investigations,
the target space was assumed to be flat. In the present paper,
we investigate the deformation of
the Dp-brane world-volume in curved background.
We choose a background such that the metric tensor $G_{\mu\nu}$ is constant,
the antisymmetric tensor $B_{\mu\nu}$ is linear in coordinate
and its field strength $B_{\mu\nu\rho}$ is nonvanishing parameter \cite{VS,CSS}.
This choice is in accordance with the space-time equations of motion,
obtained from the world-sheet conformal invariance,
as far as we can neglect the Ricci tensor.
So, we demand that $B_{\mu\nu\rho}$ is an infinitesimal parameter and
we work in the leading order in $B_{\mu\nu\rho}$ throughout the whole paper.
The Ricci tensor is thus neglected as an infinitesimal of the second order.
We call this choice the {\it{weakly curved background}}.
Physically, this case corresponds to the embedding
of a curved Dp-brane into a curved background.

The open string with non-vanishing field strength
of the Kalb-Ramond field has been investigated in Refs. \cite{CSS,HKK}.
The correlation functions have been computed on the disc,
and, therefrom, the Kontsevich product has been extracted.
The considerations in Ref. \cite{CSS} have been
restricted to the weakly curved background,
while that of Ref. \cite{HKK} has been restricted to the first order
in the derivatives of the background fields.

In \cite{HY}, the same problem has been considered using the canonical
method and some approximations based on low energy limits.
The main result is the new type of noncommutativity relation,
where the
noncommutativity parameter depends not only on coordinate but also
on momenta.
In Lagrangian formulation, it means that it depends on
coordinates' time derivatives.
This form of parameter has not been observed
by the path integral method, in Refs.\cite{CSS,HKK}.
So, the results obtained in the treatment of the same physical problem
with a different formalism (in the Refs. \cite{CSS,HKK} using path
integral method and in the Ref. \cite{HY} using canonical method)
are not the same but differ essentially.

Therefore, to be able to better elucidate the evident
discrepancies of these results, we developed a systematic canonical
approach in which these ambiguities could be solved properly and
the relation between \cite{HY} and \cite{CSS,HKK} could be
clarified. First, does the momentum dependent term exist? Second,
if this term exists under which conditions does it disappear,
like in Refs.\cite{CSS,HKK}?
Third, might there exist some new momentum
dependent terms, missed in \cite{HY} as a consequences of the low
energy limit assumption.

In the present paper the problem of the open string in the weakly
curved background is treated using canonical methods.
The approach applied to the constant background fields \cite{ACL,SN}
is generalized to the case of the curved one.
The boundary conditions are treated as canonical constraints.
Using Dirac requirement (that the time derivatives of the primary
constraints are also constraints) and Lagrangian equations
of motion we obtain the infinite set of constraints in the
Lagrangian form.

Following the line of Refs. \cite{SN} using the Taylor expansion,
we represent this infinite set of constraints at point
($\sigma=0$ and $\sigma=\pi$) with two $\sigma-$dependent constraints,
even and odd under world-sheet parity transformation
($\Omega : \sigma \rightarrow -\sigma$).
It is remarkable that these constraints can be expressed in compact form,
in terms of coordinates,
their first $\sigma$ and $\tau$ derivatives, and their integrals.

At this point and thereafter, we switched from Lagrangian to Hamiltonian method.
We checked the validity of the procedure by calculating the Poisson bracket
between Hamiltonian and the constraints,
reaching the conclusion that these
are, in fact, Hamiltonian constraints and that they
form a complete set of constraints.

All constraints except the zero modes \cite{BS} are of the second class,
and we solve them explicitly.
On this solution, the original canonical variables
can be expressed in terms of the effective ones.
Imposing $2\pi-$periodicity,
the constraints at $\sigma=\pi$ can be expressed
in terms of that at $\sigma=0$.
We separately solve symmetric and antisymmetric parts
of the constraints and express $\Omega$-odd
variables in terms of $\Omega-$even ones.
So, the constraints appear as particular orbifold conditions, reducing
the initial phase space to the $\Omega-$even and $2\pi-$periodic one.

The transition from the initial phase space to the effective
phase space on the orbifold requires some comment about the
corresponding canonical brackets.
We make a transition to the effective phase space with variables
$q^\mu$ and $p_\mu$,
$2\pi$-periodic and symmetric under the
transformation $\sigma\rightarrow -\sigma$, with
$\sigma\in[-\pi,\pi]$, in two steps.

In the initial phase space, with
variables $x^\mu$ and $\pi_\mu$,
we use the standard Poisson
brackets with $\sigma\in[0,\pi]$.
Because the basic effective canonical variables $q^\mu$ and $p_\mu$
(${\bar q}^\mu$ and ${\bar p}_\mu$)
are not arbitrary functions but contain only
the even (odd) powers in $\sigma$,
their brackets do not close on the standard $\delta$-function on
the interval $[0,\pi]$, but on the symmetric (antisymmetric)
$\delta$-function on the interval $[-\pi,\pi]$ times 2
(see Appendix \ref{sec:starb}).

We also impose boundary
condition $\Gamma_\mu=0$. Instead of using the Dirac brackets
associated with the second class constraints $\Gamma_\mu$, we
solve the constraints $\Gamma_\mu=0$ and then use the equivalent
star brackets between the variables restricted to the constrained
subspace.

The initial coordinates $x^\mu$ depend both on the effective
coordinates $q^\mu$ and their canonically conjugated momenta $p_\mu$.
This fact is a source of noncommutativity.
The coefficient in front
of the momenta $p_\mu$ is not a constant, as in the case of the flat
background, but depends on effective coordinates $q^\mu$.
Because to this, the noncommutative parameter will also depend on $q^\mu$.
This fact is the source of nonassociativity.

We want to stress that, even in the curved background,
after nontrivial calculations, it turns out
that only end-points of the string are noncommutative,
while interior of the string commutes.
At world-sheets boundary,
$\Omega-$odd parts of the coordinates vanish
($\bar{q}(0)=0$ and $\bar{q}(\pi)=0$) and consequently
the effective coordinate is equal to the original one.
So, we can say that noncommutative parameter depends on
original variable $x^\mu$. Formally, it has the same form as
in the flat background, but now Kalb-Ramond field is
infinitesimal and linear in coordinate.

\section{Open-string propagation in a weakly curved background}
\cleq
Let us consider the open bosonic string in the nontrivial background
defined by space-time fields:
the metric $G_{\mu\nu}$ and the Kalb-Ramond antisymmetric tensor $B_{\mu\nu}$.
The propagation is described by the action \cite{FTFCB,GSW}
\begin{equation}\label{eq:action}
S = \kappa \int_{\Sigma} d^2\xi \Big[\frac{1}{2}\eta^{\alpha\beta}G_{\mu\nu}(x)
+{\epsilon^{\alpha\beta}}B_{\mu\nu}(x)\Big]
\partial_{\alpha}x^{\mu}\partial_{\beta}x^{\nu},
\quad (\varepsilon^{01}=-1),
\end{equation}
where integration goes over two-dimensional world-sheet $\Sigma$
with coordinates $\xi^\alpha,\ \alpha=0,1$.
By $x^{\mu}(\xi),\ \mu=0,1,...,D-1$ we denote the coordinates of the D-dimensional space-time.
Throughout the paper we will use notation
$\xi^{0}=\tau,\ \xi^{1}=\sigma$ and
$\dot{x}=\frac{\partial x}{\partial\tau}$,
$x^\prime=\frac{\partial x}{\partial\sigma}$.

In order to preserve the quantum world sheet conformal invariance,
the $\beta$ functions for both background fields must vanish
as necessary conditions for the consistency of the theory.
To the lowest order in string slope parameter $\alpha^\prime$,
they have the form \cite{FTFCB}
\begin{equation}\label{eq:beta}
{\beta}^{G}_{\mu \nu} \equiv R_{\mu \nu} - \frac{1}{4} B_{\mu \rho \sigma}
B_{\nu}^{\ \rho \sigma}=0\, ,
\end{equation}
\begin{equation}\label{eq:beta1}
\beta^B_{\mu \nu} \equiv D_\rho B^{\rho}_{\ \mu \nu} = 0 \, .
\end{equation}
Here,
$B_{\mu\nu\rho}=\partial_\mu B_{\nu\rho}+\partial_\nu B_{\rho\mu}+\partial_\rho B_{\mu\nu}$
is the field strength of the field $B_{\mu \nu}$, and
$R_{\mu \nu}$ and $D_\mu$ are Ricci tensor and the
covariant derivative with respect to space-time metric.

In fact, to fulfill conformal invariance, it is necessary to introduce additional
background field, the dilaton field $\Phi$ and corresponding $\beta-$function.
Only derivatives of dilaton field give contribution to all $\beta-$functions, so that
the space-times equations of motion (\ref{eq:beta}) and (\ref{eq:beta1}) are correct
under the assumption $\Phi = const$.

It is an enormous task to make further progress with arbitrary background
fields. Instead, we can employ a particular solution of the space-time field
equations. We want to have the solution which admit curved background,
but to be technically as simple as possible.

It is clear that nonzero Ricci tensor $R_{\mu\nu}$
implies nontrivial $B_{\mu\nu\rho}$.
Following \cite{VS,CSS}, we choose the
field strength of the Kalb Ramond field to be constant
($B_{\mu\nu\rho}=const$) and infinitesimally small.
This solves equation  (\ref{eq:beta1}), and we can
neglect the curvature $R_{\mu\nu}$ in (\ref{eq:beta})
as an infinitesimal of the second order.
Consequently, in the leading order, the solution of the
space-time equations of the motion produces
the following background fields:
\begin{eqnarray}\label{eq:gb}
G_{\mu\nu}=const,\quad
B_{\mu\nu}[x]=\frac{1}{3}B_{\mu\nu\rho}x^\rho,
\end{eqnarray}
where the parameter $B_{\mu\nu\rho}$ is constant and infinitesimally small.
Through the paper, we will work up to its first order.
So, the chosen background is "weakly curved"
as a consequence of the infinitesimally small
Kalb-Ramond field $B_{\mu\nu}$, while
the contribution of the Ricci curvature $R_{\mu\nu}$ can be neglected.

In the case of open string,
the minimal action principle produces
the equation of motion
\begin{equation}\label{eq:motion}
{\ddot{x}}^\mu=x^{\prime\prime\mu}
-2B^\mu_{\ \nu\rho}{\dot{x}}^\nu x^{\prime\rho},
\end{equation}
and the boundary conditions on string end points
\begin{equation}\label{eq:bonc}
\gamma_{0}^\mu\Big{|}_{\sigma=0,\pi}=0,
\end{equation}
where we have introduced the variable
\begin{equation}
\gamma_{0}^\mu=x'^\mu-2(G^{-1}B)^\mu_{\ \nu}\dot{x}^\nu.
\end{equation}
Note that linear background field $B_{\mu\nu}$ contributes to equation of motion through its field strength.
This is an essential difference from the case of the constant $B_{\mu\nu}$, when it does not appear in the
equation of motion ($B_{\mu\nu\rho}=0$), and the second term in the action (\ref{eq:action}) is topological.
\section{Lagrangian consistency condition}
\cleq
We are going to treat
the boundary conditions (\ref{eq:bonc}) as the constraints.
Because they must be conserved in time,
theirs time derivative produces the new constraints,
for which we again require time conservation.
For technical reasons, instead of applying Dirac consistency procedure,
we will use analogous Lagrangian consistency procedure.
\subsection{Infinite set of constraints}
\cleq
Starting with the boundary condition $\gamma^{\mu}_{0}$ as a constraint,
with the help of the equation of motion,
we obtain the infinite set of constraints at the string end-points
\begin{equation}
\gamma_{n}^\mu\Big{|}_{\sigma=0,\pi}=0,\quad
\gamma_{n+1}^\mu\equiv\dot{\gamma}_{n}^\mu.\quad\quad\quad (n\geq 0)
\end{equation}
In order to find the explicit form of these constraints,
we introduce the following functions:
\begin{eqnarray}\label{eq:functions}
\gamma^\mu=\gamma_{0}^\mu=x'^\mu-2(G^{-1}B)^\mu_{\ \nu}\dot{x}^\nu,&&
{\tilde{\gamma}}^\mu=\dot{x}^\mu-2(G^{-1}B)^\mu_{\ \nu} x'^\nu
\nonumber\\
Q_{n}^{\alpha\beta}={\dot{x}}^{(n)\alpha}x^{(n+1)\beta},&&
R_{n}^{\alpha\beta}=x^{(n+2)\alpha}x^{(n+1)\beta}
+{\dot{x}}^{(n)\alpha}{\dot{x}}^{(n+1)\beta},
\end{eqnarray}
where $x^{(n)\alpha}=\frac{{\partial}^{n}}{\partial{\sigma}^{n}}x^\alpha$.
On the equation of motion (\ref{eq:motion}),
theirs time derivatives in the leading order are
\begin{eqnarray}
{\dot{\gamma}}^\mu={\tilde{\gamma}}^{\prime\mu},&&
{\dot{\tilde{\gamma}}}^\mu={\gamma}^{\prime\mu}-\frac{2}{3}
B^\mu_{\ \alpha\beta}Q_{0}^{\alpha\beta},\nonumber\\
{\dot{Q}}_{n}^{\alpha\beta}=R_{n}^{\alpha\beta},&&
{\dot{R}}_{n}^{\alpha\beta}=Q_{n}^{\prime\prime\alpha\beta}
-4Q_{n+1}^{\alpha\beta}.
\end{eqnarray}
Therefore, their second time derivatives are closed on the same set of functions:
\begin{eqnarray}
{\ddot{\gamma}}^\mu={\gamma}^{\prime\prime\mu}
-\frac{2}{3}B^\mu_{\ \alpha\beta}Q_{0}^{\prime\alpha\beta},&&
{\ddot{\tilde{\gamma}}}^\mu={\tilde{\gamma}}^{\prime\prime\mu}
-\frac{2}{3}B^\mu_{\ \alpha\beta}R_{0}^{\alpha\beta},\nonumber\\
{\ddot{Q}}_{n}^{\alpha\beta}=Q_{n}^{\prime\prime\alpha\beta}
-4Q_{n+1}^{\alpha\beta},&&
{\ddot{R}}_{n}^{\alpha\beta}=R_{n}^{\prime\prime\alpha\beta}
-4R_{n+1}^{\alpha\beta}.
\end{eqnarray}

It is clear that the constraints with even indices,
$\gamma_{2n}^\mu$, depend on $\gamma^\mu$ and $Q^{\alpha\beta}$,
and the ones with odd indices,
$\gamma_{2n+1}^\mu$, depend on ${\tilde{\gamma}}^\mu$ and $R^{\alpha\beta}$.
Moreover, notice that every term in $\gamma_{n}^\mu$ has exactly $n+1$ derivatives
over $\tau$ and $\sigma$. So, the expression of $\gamma_{n}^\mu$ should have
the form
\begin{eqnarray}
\gamma_{2n}^\mu&=&
\gamma^{(2n)\mu}
-\frac{2}{3}B^{\mu}_{\ \alpha\beta}
\sum_{k=0}^{n-1}\alpha^{k}_{2n}Q_{k}^{(2n-2k-1)\alpha\beta},\qquad\qquad (n\geq{1})
\nonumber\\
\gamma_{2n+1}^\mu&=&
{\tilde{\gamma}}^{(2n+1)\mu}
-\frac{2}{3}B^{\mu}_{\ \alpha\beta}
\sum_{k=0}^{n-1}\beta^{k}_{2n}R_{k}^{(2n-2k-1)\alpha\beta},\quad\qquad (n\geq{1})
\end{eqnarray}
with unknown constants $\alpha^{k}_{2n}$ and $\beta^{k}_{2n}$.
We have already seen that
$\gamma_{0}^\mu=\gamma^\mu$ and $\gamma_{1}^\mu={\tilde{\gamma}}^{\prime\mu}$.
From the definition $\gamma_{2n+2}^\mu={\ddot{\gamma}}_{2n}^\mu$,
we obtain the recurrence relation
\begin{eqnarray}
&&\alpha^{0}_{2n+2}=\alpha^{0}_{2n}+1,\nonumber\\
&&\alpha^{k}_{2n+2}=\alpha^{k}_{2n}-4\alpha^{k-1}_{2n},\qquad\qquad
({k=1,\cdots, n-1})\nonumber\\
&&\alpha^{n}_{2n+2}=-4\alpha^{n-1}_{2n},
\end{eqnarray}
with the solution
\begin{equation}
\alpha^{k}_{2n}=(-4)^{k}{n \choose {k+1}},\qquad {k=0,\cdots,n-1}\ .
\end{equation}
Using $\gamma_{2n+1}^\mu=\dot{\gamma}_{2n}^\mu$,
we conclude that $\beta^{k}_{2n}=\alpha^{k}_{2n}=(-4)^{k}{n \choose {k+1}}$.
\subsection{Compact form of the constraints at $\sigma=0$}
We obtained the explicit form of the infinite set of constraints.
Let us now multiply every constraint $\gamma_{n}^\mu\Big{|}_{\sigma=0}$
with the appropriate power of $\sigma$ and sum separately odd and even powers in
$\sigma$. In this way, we gathered the infinite set of conditions into
only two $\sigma$-dependent ones:
\begin{eqnarray}
\Gamma^\mu_{S}(\sigma)=0,&&\Gamma^\mu_{A}(\sigma)=0,
\end{eqnarray}
with
\begin{eqnarray}\label{eq:gsa}
\Gamma^\mu_{S}(\sigma)&\equiv&\sum_{n=0}^{\infty}\frac{\sigma^{2n}}{(2n)!}
\gamma_{2n}^\mu\Big{|}_{\sigma=0}
=\gamma_{S}^\mu(\sigma)
-\frac{2}{3}B^\mu_{\ \alpha\beta}\sum_{k=0}^{\infty}
(\Gamma^{Q})_{k}^{\alpha\beta}(\sigma),
\nonumber\\
\Gamma^\mu_{A}(\sigma)&\equiv&\sum_{n=0}^{\infty}\frac{\sigma^{2n+1}}{(2n+1)!}
\gamma_{2n+1}^\mu\Big{|}_{\sigma=0}
={\tilde{\gamma}}_{A}^\mu(\sigma)
-\frac{2}{3}B^\mu_{\ \alpha\beta}\sum_{k=0}^{\infty}
({\Gamma}^{R})_{k}^{\alpha\beta}(\sigma),
\nonumber\\
\end{eqnarray}
where we introduced symmetric part of $\gamma^\mu$
and antisymmetric part of ${\tilde{\gamma}}^\mu$, defined in (\ref{eq:functions}),
\begin{eqnarray}\label{eq:gammas}
\gamma_{S}^\mu(\sigma)\equiv\sum_{n=0}^{\infty}\frac{\sigma^{2n}}{(2n)!}
\gamma^{(2n)\mu}\Big{|}_{\sigma=0},
&&
{\tilde{\gamma}}_{A}^\mu(\sigma)\equiv\sum_{n=0}^{\infty}\frac{\sigma^{2n+1}}{(2n+1)!}
{\tilde{\gamma}}^{(2n+1)\mu}\Big{|}_{\sigma=0}
\end{eqnarray}
and
\begin{eqnarray}\label{eq:gsum}
(\Gamma^{Q})_{k}^{\alpha\beta}(\sigma)&\equiv&
\sum_{n=k+1}^{\infty}(-4)^{k}{n \choose {k+1}}
\frac{\sigma^{2n}}{(2n)!}
Q_{k}^{(2n-2k-1)\alpha\beta}\Big{|}_{\sigma=0},
\nonumber\\
({\Gamma}^{R})_{k}^{\alpha\beta}(\sigma)&\equiv&
\sum_{n=k+1}^{\infty}(-4)^{k}{n \choose {k+1}}
\frac{\sigma^{2n+1}}{(2n+1)!}
R_{k}^{(2n-2k-1)\alpha\beta}\Big{|}_{\sigma=0}.
\nonumber\\
\end{eqnarray}
These sums can be represented in the integral form
(see Appendix \ref{sec:ioform})
\begin{eqnarray}
(\Gamma^{Q})_{k}^{\ \alpha\beta}(\sigma)&=&
\frac{(-1)^{k}\sigma}{2(k+1)!}
\int_{0}^{\sigma}d\sigma_{1}^{2}
\int_{0}^{\sigma_{1}}d\sigma_{2}^{2}\cdots
\int_{0}^{\sigma_{k-1}}d\sigma_{k}^{2}
(Q_{A})_{k}^{\alpha\beta}(\sigma_{k}),
\end{eqnarray}
in terms of the antisymmetric part of $Q_{k}^{\alpha\beta}$
\begin{eqnarray}
(Q_{A})_{k}^{\alpha\beta}(\sigma)&\equiv&
\sum_{n=0}^{\infty}\frac{{\sigma}^{2n+1}}{(2n+1)!}
Q_{k}^{(2n+1)\alpha\beta}\Big{|}_{0}=
[{\dot{x}}^{(k)\alpha}x^{(k+1)\beta}]_{A}.
\end{eqnarray}
Notice that
\begin{equation}
(\Gamma^{R})_{k}^{\prime\alpha\beta}(\sigma)=
(\Gamma^{Q})_{k}^{\ \alpha\beta}(\sigma)\Big{|}_{Q\mapsto R},
\end{equation}
so that, by analogy, we can write
\begin{eqnarray}\label{eq:r}
(\Gamma^{R})_{k}^{\ \alpha\beta}(\sigma)&=&
\frac{(-1)^{k}}{4(k+1)!}
\int_{0}^{\sigma}d{\sigma_{0}}^{2}
\int_{0}^{\sigma_{0}}d\sigma_{1}^{2}\cdots
\int_{0}^{\sigma_{k-1}}d{\sigma_{k}}^{2}
(R_{A})_{k}^{\alpha\beta}(\sigma_{k}),
\end{eqnarray}
in terms of the antisymmetric part of $R^{\alpha\beta}_{k}$,
\begin{eqnarray}
(R_{A})_{k}^{\alpha\beta}(\sigma)&\equiv&
\sum_{n=0}^{\infty}\frac{{\sigma}^{2n+1}}{(2n+1)!}
R_{k}^{(2n+1)\alpha\beta}\Big{|}_{0}
=[x^{(n+2)\alpha}x^{(n+1)\beta}
+{\dot{x}}^{(n)\alpha}{\dot{x}}^{(n+1)\beta}]_{A}.
\end{eqnarray}
Consequently, we can express $\Gamma^\mu_{S}(\sigma)$,
defined in (\ref{eq:gsa}), in terms of
the symmetric part of $\gamma^\mu$ and
the antisymmetric parts of
${\tilde{\gamma}}^\mu$,
$Q_{k}^{\alpha\beta}$, and
$R_{k}^{\alpha\beta}$,
defined in (\ref{eq:functions}).

In order to separate the symmetric and antisymmetric parts under $\sigma-$parity,
we introduce the new variables
\begin{eqnarray}\label{eq:qqbar}
q^\mu(\sigma)=
\sum_{n=0}^{\infty}\frac{{\sigma}^{2n}}{(2n)!}x^{(2n)\mu}\Big{|}_{\sigma=0},&&
{\bar{q}}^\mu(\sigma)=
\sum_{n=0}^{\infty}\frac{{\sigma}^{2n+1}}{(2n+1)!}x^{(2n+1)\mu}\Big{|}_{\sigma=0},
\end{eqnarray}
which we will call open-string variables.

In terms of the new variables, we have
\begin{eqnarray}\label{eq:ggqr1}
\gamma_{S}^\mu&=&{\bar{q}}^{\prime\mu}
-\frac{2}{3}B^{\mu}_{\ \nu\rho}({\dot{q}}^\nu q^\rho
+{\dot{\bar{q}}}^\nu {\bar{q}}^\rho),
\nonumber\\
{\tilde{\gamma}}_{A}^\mu&=&\dot{\bar{q}}^\mu
-\frac{2}{3}B^{\mu}_{\ \nu\rho}
(q^{\prime\nu} q^\rho+ {\bar{q}}^{\prime\nu} {\bar{q}}^\rho),
\nonumber\\
(Q_{A})_{k}^{\alpha\beta}&=&
{\dot{q}}^{(k)\alpha} q^{(k+1)\beta}+
{\dot{\bar{q}}}^{(k)\alpha}{\bar{q}}^{(k+1)\beta},
\nonumber\\
(R_{A})_{k}^{\alpha\beta}&=&
q^{(k+2)\alpha}q^{(k+1)\beta}
+{\bar{q}}^{(k+2)\alpha}{\bar{q}}^{(k+1)\beta}
\nonumber\\
&+&
{\dot{q}}^{(k)\alpha} {\dot{q}}^{(k+1)\beta}+
{\dot{\bar{q}}}^{(k)\alpha}{\dot{\bar{q}}}^{(k+1)\beta}.
\end{eqnarray}

Using two previous equations,
we can rewrite the last terms in (\ref{eq:gsa}) as
\begin{eqnarray}\label{eq:sqr}
\sum_{k=0}^{\infty}(\Gamma^{Q})_{k}^{\alpha\beta}&=&
h^{\alpha\beta}[\dot{q},q]+h^{\alpha\beta}[\dot{\bar{q}},{\bar{q}}],
\nonumber\\
\sum_{k=0}^{\infty}(\Gamma^{R})_{k}^{\alpha\beta}&=&
\int_{0}^{\sigma}d\sigma_{0}
\Big{[}h^{\alpha\beta}[q^{\prime\prime},q]+h^{\alpha\beta}[{\bar{q}}^{\prime\prime},{\bar{q}}]
\nonumber\\
&+&
h^{\alpha\beta}[\dot{q},\dot{q}]+h^{\alpha\beta}[\dot{\bar{q}},\dot{\bar{q}}]\Big{]}(\sigma_{0}),
\end{eqnarray}
where we introduced function $h^{\alpha\beta}[a,b]$
\begin{equation}\label{eq:h1}
h^{\alpha\beta}[a,b](\sigma)=
\frac{\sigma}{2}
\sum_{k=0}^{\infty}\frac{(-1)^{k}}{(k+1)!}
\int_{0}^{\sigma}d\sigma_{1}^{2}\cdots
\int_{0}^{\sigma_{k-1}}d\sigma_{k}^{2}
a^{(k)\alpha}(\sigma_{k}) b^{(k+1)\beta}(\sigma_{k}).
\end{equation}
Using (\ref{eq:h}), (\ref{eq:hn}),
we rewrite constraints (\ref{eq:gsa})  as
\begin{eqnarray}\label{eq:gsa1}
\Gamma^{\mu}_{S}(\sigma)&=&
{\bar{q}}^{\prime\mu}-\frac{2}{3}B^{\mu}_{\ \nu\rho}
[{\dot{q}}^\nu q^\rho
+\frac{1}{2}{\dot{Q}}^\nu q^{\prime\rho}
+\frac{3}{2}{\dot{\bar{q}}}^\nu {\bar{q}}^\rho],
\nonumber\\
\Gamma^{\mu}_{A}(\sigma)&=&
{\dot{\bar{q}}}^\mu
-\frac{2}{3}B^{\mu}_{\ \nu\rho}
[q^{\prime\nu}q^\rho
+\frac{1}{2}{\dot{Q}}^\nu{\dot{q}}^\rho
+\frac{3}{2}{\bar{q}}^{\prime\nu}{\bar{q}}^\rho].
\end{eqnarray}
\section{Canonical form of the constraints at $\sigma=0$}
\cleq
Now, we are ready to make transition from Lagrangian to Hamiltonian approach.
Let us first introduce the canonical momenta corresponding to coordinates $x^\mu$,
\begin{equation}
\pi_\mu=\kappa(G_{\mu\nu}{\dot{x}}^\nu-2B_{\mu\nu}x'^\nu),
\end{equation}
and the canonical Hamiltonian
\begin{equation}\label{eq:hc}
H_{C}=\int_{0}^{\pi}d\sigma
\Big{[}\frac{1}{2\kappa}(G^{-1})^{\mu\nu}\pi_\mu \pi_\nu
+\frac{\kappa}{2}G_{\mu\nu}x^{\prime\mu}x^{\prime\nu}
+\frac{2}{3}B^{\mu}_{\ \nu\rho}\pi_\mu x^{\prime\nu}x^{\rho}
\Big{]}.
\end{equation}
Similarly, as in (\ref{eq:qqbar}), we introduce new, open-string momenta
\begin{eqnarray}\label{eq:qp}
p^\mu(\sigma)=
\sum_{n=0}^{\infty}\frac{{\sigma}^{2n}}{(2n)!}\pi^{(2n)\mu}\Big{|}_{\sigma=0},&&
\bar{p}^\mu(\sigma)=
\sum_{n=0}^{\infty}\frac{{\sigma}^{2n+1}}{(2n+1)!}\pi^{(2n+1)\mu}\Big{|}_{\sigma=0},
\end{eqnarray}
and rewrite the constraints (\ref{eq:gsa1}) in a canonical form
\begin{eqnarray}\label{eq:g}
\Gamma^\mu_{S}(\sigma)&=&
{\bar{q}}^{\prime\mu}+\theta^{\mu\nu}(q)p_\nu
+\frac{1}{2}\theta^{\prime\mu\nu}(q)P_\nu
+\frac{3}{2}\theta^{\mu\nu}(\bar{q}){\bar{p}}_\nu,
\nonumber\\
\kappa\Gamma^\mu_{A}(\sigma)&=&
(G^{-1})^{\mu\nu}{\bar{p}}_\nu
+\frac{\kappa^{2}}{2}\theta^{\mu}_{\ \nu}(\bar{q}){\bar{q}}^{\prime\nu}
+\frac{1}{2}\theta^{\mu\nu}(p)P_\nu,
\end{eqnarray}
where
\begin{eqnarray}\label{eq:teta}
\theta^{\mu\nu}[q(\sigma)]&\equiv&-\frac{2}{3\kappa}B^{\mu\nu}_{\ \ \rho}q^\rho(\sigma)
=-\frac{2}{\kappa}(G^{-1})^{\mu\alpha}B_{\alpha\beta}[q(\sigma)](G^{-1})^{\beta\nu},
\nonumber\\
P_\mu(\sigma)&\equiv&\int_{0}^{\sigma}d\eta p_\mu(\eta).
\end{eqnarray}

Note that, from the standard Poisson brackets
\begin{equation}\label{eq:poib}
\{x^\mu(\sigma),\pi_\nu(\bar{\sigma})\}=\delta^{\mu}_{\ \nu}\delta(\sigma-\bar{\sigma}),
\end{equation}
we have two non-trivial relations for $\Omega$ even and odd subspaces
\begin{eqnarray}\label{eq:poisson}
\{q^\mu(\sigma),p_\nu(\bar{\sigma})\}=2\delta^{\mu}_{\
\nu}\delta_{S}(\sigma,\bar{\sigma}),&&
\{{\bar{q}}^\mu(\sigma),{\bar{p}}_\nu(\bar{\sigma})\}=2\delta^{\mu}_{\
\nu}\delta_{A}(\sigma,\bar{\sigma}),
\end{eqnarray}
where $\delta_{S}$ and $\delta_{A}$ are defined in (\ref{eq:dsa}).
Because $\Gamma_{S}^\mu$ and $\Gamma_{A}^\mu$,
as the symmetric and antisymmetric functions, are independent,
it is enough to consider the constraint
$\Gamma^\mu=-\kappa(\Gamma_{S}^\mu-\Gamma_{A}^\mu)$.
It weakly commutes with the Hamiltonian
\begin{equation}
\{ H_c, \Gamma^\mu (\sigma) \} = \Gamma^{\prime\mu} (\sigma),
\end{equation}
and, therefore, there are no more constraints.
We can calculate the Poisson brackets,
up to the term linear in small parameter $B^{\mu\nu\rho}$, as
\begin{eqnarray}\label{eq:vg}
\{\Gamma^\mu(\sigma),\Gamma^\nu(\bar{\sigma})\}&=&
-\kappa(G^{-1})^{\mu\nu}\delta^{\prime}(\sigma-\bar{\sigma})
\nonumber\\
&-&B^{\mu\nu\rho}[{\bar{p}}_\rho(\sigma)-\kappa G_{\rho\tau}{\bar{q}}^{\prime\tau}(\sigma)]
\delta(\sigma-\bar{\sigma})
\nonumber\\
&\approx&
-\kappa(G^{-1})^{\mu\nu}\delta^{\prime}(\sigma-\bar{\sigma}).
\end{eqnarray}
The sign $\approx$ is a weak equality which,
in the canonical approach, means equality on the constraints.
In particular case, with the help of (\ref{eq:g}),
from $\Gamma_{S}^\mu\approx 0$ and $\Gamma_{A}^\mu\approx 0$,
it follows that ${\bar{q}}^\mu$ and ${\bar{p}}_\nu$ are
proportional to $B^{\mu\nu\rho}$, so that the term in front
of $\delta(\sigma-\bar{\sigma})$ in (\ref{eq:vg}) is
infinitesimal of the second order.
Therefore, we conclude that $\Gamma^\mu$ and, consequently,
$\Gamma_{S}^\mu$ and $\Gamma_{A}^\mu$
are the second class constraints.
We will look for their solution in Sec. \ref{sec:sol}.

There is a slight improvement of the above conclusion.
The Poisson bracket between constraints $\Gamma^\mu$ is closed on
$\delta^\prime(\sigma-\bar{\sigma})$ and
not on $\delta(\sigma-\bar{\sigma})$
function. So, the zero mode of $\Gamma^\mu(\sigma)$,
\begin{equation}\label{eq:gamanula}
\Gamma^\mu_{0}=\int_{0}^{\pi}d\sigma \Gamma^\mu(\sigma),
\end{equation}
is the first class constraint,
because $\{\Gamma^\mu_{0},\Gamma^\nu(\sigma)\}=0$.
Consequently,
it is a generator of gauge symmetry with constant parameter.
We will use this fact at the end of Sec. \ref{sec:sol} to gauge away
the center of mass of the coordinate.
\section{Constraints at $\sigma=\pi$}
\cleq
In order to derive constraints at the other string endpoint $\sigma=\pi$,
we will multiply every constraint $\gamma_{n}^\mu\Big{|}_{\sigma=\pi}$
with the appropriate power of $\sigma-\pi$ and sum separately odd and
even powers in $\sigma-\pi$.
We obtain two new $\sigma$-dependent constraints:
\begin{eqnarray}
{\bar{\Gamma}}^\mu_{S}(\sigma)=0,&&{\bar{\Gamma}}^\mu_{A}(\sigma)=0,
\end{eqnarray}
where
\begin{eqnarray}\label{eq:bgsa}
{\bar{\Gamma}}^\mu_{S}(\sigma)&=&\sum_{n=0}^{\infty}\frac{(\sigma-\pi)^{2n}}{(2n)!}
\gamma_{2n}^\mu\Big{|}_{\sigma=\pi}
={\bar{\gamma}}_{S}^\mu(\sigma)
-\frac{2}{3}B^\mu_{\ \alpha\beta}\sum_{k=0}^{\infty}
({\bar{\Gamma}}^{Q})_{k}^{\alpha\beta}(\sigma),
\nonumber\\
{\bar{\Gamma}}^\mu_{A}(\sigma)&=&\sum_{n=0}^{\infty}\frac{(\sigma-\pi)^{2n+1}}{(2n+1)!}
{\gamma}_{2n+1}^\mu\Big{|}_{\sigma=\pi}
={\bar{{\tilde{\gamma}}}}_{A}^\mu(\sigma)
-\frac{2}{3}B^\mu_{\ \alpha\beta}\sum_{k=0}^{\infty}
({{\bar{\Gamma}}}^{R})_{k}^{\alpha\beta}(\sigma),
\nonumber\\
\end{eqnarray}
and
\begin{eqnarray}
{\bar{\gamma}}_{S}^\mu(\sigma)=\sum_{n=0}^{\infty}\frac{(\sigma-\pi)^{2n}}{(2n)!}
\gamma^{(2n)\mu}\Big{|}_{\sigma=\pi},
&&
{\bar{\tilde{\gamma}}}_{A}^\mu(\sigma)=\sum_{n=0}^{\infty}\frac{(\sigma-\pi)^{2n+1}}{(2n+1)!}
{\tilde{\gamma}}^{(2n+1)\mu}\Big{|}_{\sigma=\pi},
\end{eqnarray}
\begin{eqnarray}\label{eq:gqnad}
({\bar{\Gamma}}^{Q})_{k}^{\alpha\beta}(\sigma)&=&
\sum_{n=k+1}^{\infty}(-4)^{k}{n \choose {k+1}}
\frac{(\sigma-\pi)^{2n}}{(2n)!}
Q_{k}^{(2n-2k-1)\alpha\beta}\Big{|}_{\sigma=\pi},
\nonumber\\
({\bar{\Gamma}}^{R})_{k}^{\alpha\beta}(\sigma)&=&
\sum_{n=k+1}^{\infty}(-4)^{k}{n \choose {k+1}}
\frac{(\sigma-\pi)^{2n+1}}{(2n+1)!}
R_{k}^{(2n-2k-1)\alpha\beta}\Big{|}_{\sigma=\pi}.
\nonumber\\
\end{eqnarray}

The $({\bar{\Gamma}}^{Q})_{k}^{\alpha\beta}(\sigma)$
can be written in the integral form (Appendix \ref{sec:ipform})
\begin{equation}\label{eq:bgq1}
({\bar{\Gamma}}^{Q})_{k}^{\ \alpha\beta}(\sigma)=
\frac{\sigma-\pi}{2(k+1)!}
\int_{\sigma}^{\pi}d(\sigma_{1}-\pi)^{2}
\int_{\sigma_{1}}^{\pi}d(\sigma_{2}-\pi)^{2}\cdots
\int_{\sigma_{k-1}}^{\pi}d(\sigma_{k}-\pi)^{2}
({\bar{Q}}_{A})_{k}^{\alpha\beta}(\sigma_{k}),
\end{equation}
with
\begin{equation}
({\bar{Q}}_{A})^{\alpha\beta}_{k}(\sigma)\equiv\sum_{n=0}^{\infty}\frac{(\sigma-\pi)^{2n+1}}{(2n+1)!}
{Q_{k}}^{(2n+1)\alpha\beta}\Big{|}_{\sigma=\pi}.
\end{equation}

In analogy with (\ref{eq:qqbar}) and (\ref{eq:qp}), we introduce
symmetric and antisymmetric variables in the neighborhood of $\sigma=\pi$,
\begin{eqnarray}\label{eq:qp1}
{\tilde{q}}^\mu(\sigma)=
\sum_{n=0}^{\infty}\frac{{\sigma}^{2n}}{(2n)!}x^{(2n)\mu}\Big{|}_{\sigma=\pi},&&
{\bar{\tilde{q}}}^\mu(\sigma)=
-\sum_{n=0}^{\infty}\frac{{\sigma}^{2n+1}}{(2n+1)!}x^{(2n+1)\mu}\Big{|}_{\sigma=\pi},
\nonumber\\
{\tilde{p}}^\mu(\sigma)=
\sum_{n=0}^{\infty}\frac{{\sigma}^{2n}}{(2n)!}\pi^{(2n)\mu}\Big{|}_{\sigma=\pi},&&
{\bar{\tilde{p}}}^\mu(\sigma)=
-\sum_{n=0}^{\infty}\frac{{\sigma}^{2n+1}}{(2n+1)!}\pi^{(2n+1)\mu}\Big{|}_{\sigma=\pi}.
\nonumber\\
\end{eqnarray}

In the canonical form,
in terms of the new variables, Eq.
(\ref{eq:bgq1}) can be rewritten as
\begin{eqnarray}
({\bar{\Gamma}}^{Q})_{k}^{\ \alpha\beta}(\sigma)&=&
\frac{1}{\kappa}(G^{-1})^{\alpha\gamma}
\frac{\pi-\sigma}{2(k+1)!}(-1)^{k}
\nonumber\\
&\cdot&
\int_{0}^{\pi-\sigma}d{\eta_{1}}^{2}
\cdots
\int_{0}^{\eta_{k-1}}d{\eta_{k}}^{2}
\Big{[}{\tilde{p}}^{(k)}_\gamma {\tilde{q}}^{(k+1)\beta}+
{\tilde{{\bar{p}}}}^{(k)}_\gamma {\tilde{{\bar{q}}}}^{(k+1)\beta}\Big{]}(\eta_{k}),
\end{eqnarray}
where we introduced $\eta_{k}=\pi-\sigma_{k}$.
As before, we can observe that
\begin{equation}
({\bar{\Gamma}}^{R})_{k}^{\prime\alpha\beta}(\sigma)=
({\bar{\Gamma}}^{Q})_{k}^{\ \alpha\beta}(\sigma)\Big{|}_{Q\mapsto R},
\end{equation}
stands.
Finally, we can write the explicit form of
the $\sigma$-dependent constraints in $\sigma=\pi$:
\begin{eqnarray}\label{eq:bgp}
{\bar{\Gamma}}^\mu_{S}(\sigma)&=&\Big{\{}
-{\bar{\tilde{q}}}^{\prime\mu}
+\theta^{\mu\nu}[\tilde{q}]{\tilde{p}}_{\nu}
+\frac{1}{2}\theta^{\prime\mu\nu}[\tilde{q}]{\tilde{P}}_\nu
+\frac{3}{2}\theta^{\mu\nu}[\bar{\tilde{q}}]{\bar{\tilde{p}}}_\nu
\Big{\}}(\pi-\sigma),
\nonumber\\
\kappa{\bar{\Gamma}}^\mu_{A}(\sigma)&=&\Big{\{}(G^{-1})^{\mu\nu}{\bar{\tilde{p}}}_\nu
-\frac{\kappa^{2}}{2}\theta^\mu_{\ \nu}[\bar{\tilde{q}}]{\bar{\tilde{q}}}^{\prime\nu}
-\frac{1}{2}\theta^{\mu\nu}[\tilde{p}]{\tilde{P}}_\nu
\Big{\}}(\pi-\sigma),
\end{eqnarray}
where all variables on the right-hand side depend on $\pi-\sigma$.
Comparing (\ref{eq:g}) with (\ref{eq:bgp}),
we find the relations
\begin{eqnarray}
{\Gamma}^\mu_{S}[q,p,\bar{q},\bar{p},q^\prime,{\bar{q}}^\prime,P](\sigma)
&=&{\bar{\Gamma}}^\mu_{S}[\tilde{q},\tilde{p},\tilde{\bar{q}},
\tilde{\bar{p}},-{\tilde{q}}^\prime,-{\tilde{\bar{q}}}^\prime,-\tilde{P}](\pi-\sigma),
\nonumber\\
{\Gamma}^\mu_{A}[p,\bar{q},\bar{p},{\bar{q}}^\prime,P](\sigma)&=&
{\Gamma}^\mu_{A}[\tilde{p},\tilde{\bar{q}},\tilde{\bar{p}},-\tilde{{\bar{q}}}^\prime,-\tilde{P}](\pi-\sigma).
\end{eqnarray}

Note that, for all variables, we have
$z(\sigma)={\tilde{z}}(\pi-\sigma)$,
where $z=\{q,p,\bar{q},\bar{p}\}$.
For the corresponding $\sigma$-derivatives
and $\sigma$-integrals,
there is an additional minus
sign (e.g. $q^{\prime\mu}(\sigma)=
-{\tilde{q}}^{\prime\mu}(\pi-\sigma)$,
$P_\mu(\sigma)=-P_\mu(\pi-\sigma)$),
which is equivalent to the above relations.

With the help of (\ref{eq:qqbar}), (\ref{eq:qp}),
and (\ref{eq:qp1}) we can con\-clude that,
if we demand $2\pi$-perio\-dicity of the original coordi\-nates and momenta,
\begin{equation}\label{eq:dvapi}
x(\sigma)=x(\sigma+2\pi),\qquad \pi(\sigma)=\pi(\sigma+2\pi),
\end{equation}
by solving the $\sigma-$dependent constraints at $\sigma=0$, we solve
the $\sigma$-dependent constraints at $\sigma=\pi$ also.
\section{Noncommutativity on the string end points}\label{sec:sol}
\cleq
Instead of constructing the Dirac brackets,
we are going to solve
the second class constraints $\Gamma^\mu_{S}(\sigma)=0$ and
$\Gamma^\mu_{A}(\sigma)=0$ explicitly.
Up to the term linear in infinitesimal parameter $B_{\mu\nu\rho}$, we obtain
\begin{eqnarray}
{\bar{q}}^\mu(\sigma)&=&
-\int_{0}^{\sigma}d\sigma_{0}
\Big{(}\theta^{\mu\nu}[q]p_\nu +\frac{1}{2}\theta^{\prime\mu\nu}[q]P_\nu\Big{)}(\sigma_{0}),
\nonumber\\
{\bar{p}}_\mu(\sigma)&=&
-\frac{1}{2}\theta_{\mu}^{\ \nu}[p(\sigma)]P_\nu(\sigma).
\end{eqnarray}\label{eq:qbar}
Notice that both ${\bar{q}}^\mu$ and ${\bar{p}}_\mu$
are proportional to $B_{\mu\nu\rho}$.
Therefore, we neglected $\theta^{\mu\nu}(\bar{q}){\bar{p}}_\nu$ in $\Gamma^\mu_{S}$
and $\theta^{\mu}_{\ \nu}(\bar{q}){\bar{q}}^{\prime\nu}$ in $\Gamma^\mu_{A}$,
because they are of higher order in $B_{\mu\nu\rho}$.
By solving the constraints, we obtained the expressions for
antisymmetric variables ${\bar{q}}^\mu,\ {\bar{p}}_\mu$,
in terms of the symmetric ones, ${q}^\mu,\ {p}_\mu$.
So, we can express original variables in terms of the
new ones:
\begin{eqnarray}\label{eq:solution}
x^\mu(\sigma)&=&q^\mu(\sigma)
-\int_{0}^{\sigma}d\sigma_{0}
\Big{(}
\theta^{\mu\nu}[q]p_\nu
+\frac{1}{2}\theta^{\prime\mu\nu}[q]P_\nu
\Big{)}(\sigma_{0}),
\nonumber\\
\pi_\mu(\sigma)&=& p_\mu(\sigma)
-\frac{1}{2}\theta_{\mu}^{\ \nu}[p(\sigma)]P_\nu(\sigma).
\end{eqnarray}
Let us stress that, from the moment we solved the constraints,
the  open-string variables $q^\mu$ and $p_\nu$ become the
fundamental quantities, while the closed string variables $x^\mu$
and $\pi_\mu$ become derived ones.
So, the phase space of the effective theory
(obtained on the solution of the boundary
condition) is a subspace containing only the even powers in
$\sigma$, with the canonical variables $q^\mu$ and $p_\mu$ and star
brackets (see Appendix \ref{sec:starb})
\begin{equation}\label{eq:stbrac}
{}^{\star}\{q^\mu(\sigma),p_\nu(\bar{\sigma})\}
=2\delta^{\mu}_{\nu}\delta_{S}(\sigma,\bar{\sigma}).
\end{equation}

Since the coordinates $x^\mu$ depend both on effective coordinates $q^\mu$
and effective momenta $p_\mu$, they are noncommutative.

Using these relations, we can calculate the star bracket between composed variables $x^\mu$.
The second term in the first relation in (\ref{eq:solution})
is infinitesimal and therefore only the star brackets between
the first and second terms
give the nontrivial contribution
\begin{equation}\label{eq:qqbar6}
{}^{\star}\{q^\mu(\sigma),{\bar{q}}^\nu(\bar{\sigma})\}
=2\theta^{\mu\nu}[q(\sigma)]\theta_{S}(\bar{\sigma},\sigma)
+\int_{0}^{\bar{\sigma}}d\sigma_{0}\theta^{\prime\mu\nu}[q(\sigma_{0})]
\theta_{S}(\sigma_{0},\sigma).
\end{equation}
Substituting this result into expression for
${}^{\star}\{x^\mu(\sigma),x^\nu(\bar{\sigma})\}$ and using the properties of $\theta$
function (see Appendix \ref{sec:deltateta}), we get
\begin{equation}
{}^{\star}\{x^\mu(\sigma),x^\nu(\bar{\sigma})\}=
\Big{\{}
\theta^{\mu\nu}[q(\sigma)]+\theta^{\mu\nu}[q(\bar{\sigma})]
\Big{\}}
\theta(\sigma+\bar{\sigma}).
\end{equation}
Notice that term with $\theta(\sigma-\bar{\sigma})$ disappears.

If we separate a center-of-mass variable
$x^{\mu}_{cm}=\frac{1}{\pi}\int_{0}^{\pi}d\sigma x^\mu(\sigma)$,
we can write
\begin{equation}\label{eq:xcm}
x^\mu(\sigma)=X^\mu(\sigma)+x^{\mu}_{cm},
\end{equation}
and obtain
\begin{eqnarray}\label{eq:xx}
{}^{\star}\{X^\mu(\sigma),X^\nu(\bar{\sigma})\}
&=&\theta^{\mu\nu}[q(\sigma)]
\left\{ \begin{array}{rcl}
-1 && \sigma,\bar{\sigma}=0\\
1 && \sigma,\bar{\sigma}=\pi\\
0 && otherwise
\end{array}\right..
\end{eqnarray}
So, the interior of the string is commutative, and only the
string end-points are noncommutative. The noncommutative parameter
$\theta^{\mu\nu}$ now depends on effective coordinate $q^\mu$.
Because $\bar{q}(0)=0$ and $\bar{q}(\pi)=0$,
we can rewrite the right-hand side of (\ref{eq:xx}) in terms of $x^\mu$,
instead in terms of $q^\mu$.
In order to close algebra on the same variable $X^\mu$ using the gauge
symmetry, generated by zero mode of the constraints
$\Gamma^\mu_{0}$ (\ref{eq:gamanula}),
we can gauge away $x^\mu_{cm}$.
Therefore, the final form of noncommutativity relation takes the form
\begin{eqnarray}\label{eq:xxx}
{}^{\star}\{X^\mu(\sigma),X^\nu(\bar{\sigma})\}
&=&\theta^{\mu\nu}[X(\sigma)]
\left\{ \begin{array}{rcl}
-1 && \sigma,\bar{\sigma}=0\\
1 && \sigma,\bar{\sigma}=\pi\\
0 && otherwise
\end{array}\right..
\end{eqnarray}
Formally,
the noncommutativity parameter $\theta^{\mu\nu}$,
defined in (\ref{eq:teta}),
has the same structure as in the flat case.
But, in the curved background,
it is infinitesimally small and linear in coordinate,
as well as Kalb-Ramond field $B_{\mu\nu}$.
\section{Canonical quantization and the Kontsevich star product}
\cleq
In the quantization procedure,
we associate a corresponding operator with every variable,
and the star brackets are replaced by the commutator.
It follows from (\ref{eq:xxx}) that the noncommutativity appears
only on the string end points.
The noncommutativity parameters at $\sigma=0$ and $\sigma=\pi$
differ only in sign. So, it is enough to consider
$\sigma=0$ case
\begin{eqnarray}\label{eq:com}
[{\hat{X}}^\mu,{\hat{X}}^\nu]= - i\theta^{\mu\nu}(\hat{X}),
\end{eqnarray}
where, from now on,
we use the notation $X^\mu\equiv X^\mu(\sigma=0)$.

We are interested in the algebra of the functions
defined on $Dp$-brane world volume.
We will show that it is deformed because $Dp$ brane
propagates in a background with
the nontrivial Kalb-Rammond field $B_{\mu\nu}$.
In order to uniquely assign an operator
$\hat{f}(\hat{X})$ to any function $f(X)$,
we introduce
the Weyl prescription procedure
\begin{equation}\label{eq:weyl}
f(X)=\frac{1}{(2\pi)^{D}}\int d^{D}k \tilde{f}(k)e^{-ikX}\Rightarrow
\hat{f}(\hat{X})=\frac{1}{(2\pi)^{D}}\int d^{D}k \tilde{f}(k)e^{-ik\hat{X}}.
\end{equation}
If we have two functions and two associated operators
$f{\rightarrow}{\hat{f}},\quad g{\rightarrow}{\hat{g}}$,
we define the star product demanding the prescription
$
f\star g\rightarrow {\hat{f}}{\hat{g}}.
$
Because of the $X$ dependence of the $\theta^{\mu\nu}$,
the $\star$ is the Kontsevich product \cite{K} which,
up to the second order in $\theta^{\mu\nu}$, is equal to
\begin{eqnarray}
f\star g &=& fg +\frac{i}{2}\theta^{\mu\nu}\partial_\mu f \partial_\nu g
-\frac{1}{8}\theta^{\mu\nu}\theta^{\rho\sigma}\partial_\mu \partial_\rho f\partial_\nu\partial_\sigma g
\nonumber\\
&-&\frac{1}{12}\theta^{\rho\sigma}\partial_\sigma\theta^{\mu\nu}(\partial_\rho{\partial_\mu}f {\partial_\nu}g-
{\partial_\mu}f \partial_\rho{\partial_\nu}g)+ {\cal{O}}(\theta^{3}).
\end{eqnarray}
It can be shown that
\begin{equation}\label{eq:asoc}
(f\star g)\star h-f\star(g\star h)
=\frac{1}{6}[\theta^{\mu\sigma}\partial_\sigma\theta^{\nu\rho}
+\theta^{\nu\sigma}\partial_\sigma\theta^{\mu\rho}
+\theta^{\rho\sigma}\partial_\sigma\theta^{\mu\nu}]
\partial_\mu f \partial_\nu g \partial_\rho h + {\cal{O}}(\theta^{3}).
\end{equation}
If we denote the invers of $\theta^{\mu\nu}$ by $\theta_{\rho\sigma}$,
($\theta^{\mu\nu}\theta_{\nu\rho}=\delta^\mu_\rho$),
we can rewrite (\ref{eq:asoc}) as
\begin{eqnarray}
(f\star g)\star h-f\star(g\star h)&=&
\frac{1}{6}\theta^{\mu\alpha}\theta^{\nu\beta}\theta^{\rho\gamma}
\theta_{\alpha\beta\gamma}\partial_\mu f
\partial_\nu g \partial_\rho h,
\end{eqnarray}
where
\begin{equation}
\theta_{\mu\nu\rho}=\partial_\mu \theta_{\nu\rho}+\partial_\nu \theta_{\rho\mu}+\partial_\rho \theta_{\mu\nu},
\end{equation}
is a nonassociativity parameter.

In our case, the noncommutativity parameter
\begin{eqnarray}\label{eq:com1}
\theta^{\mu\nu}(\hat{X})\equiv - \frac{2}{3\kappa}B^{\mu\nu}_{\ \ \rho}{\hat{X}}^\rho,
\end{eqnarray}
is infinitesimal, and the Kontsevich product is associative,
because the right-hand side
of (\ref{eq:asoc}) is of the second order
in the small parameter $B_{\mu\nu\rho}$.

\section{Conclusions and discussions}
\cleq
In the present paper, we investigated the geometry of $Dp$-branes  in curved background.
We chose the simplest possible case of infinitesimally curved background,
where Kalb-Ramond field is infinitesimally small and linear in coordinate.
In such a case, we avoided working with a nonconstant metric tensor, because,
as a consequence of space-time field equations,
the Ricci tensor is an infinitesimal of the second order
and can be neglected in the leading order.

In obtaining Poisson bracket between original coordinates $x^\mu$,
it was useful to express them in terms of effective coordinates $q^\mu$
and corresponding canonical momenta $p_\mu$.
On the other hand,
Lagrangian formalism is more appropriate for working with an infinite set of constraints.
So, we used "adopted canonical approach".

We treated boundary conditions as constraints.
The basic technical problem was the derivation
of the Dirac consistency conditions. Instead to commute
the constraints with the Hamiltonian in order to obtain new constraints,
we found it more appropriate to use Lagrangian
approach.
With the help of the Lagrangian equations of motion,
we obtained the time derivatives of the primary constraints in leading orders.
According to the Dirac requirement,
they were constraints, also. Therefore,
by further application of this procedure,
we obtained the infinite set of constraints in
Lagrangian form.

Following the procedure of Refs.\cite{SN},
we substituted an infinite set of constraints at string end-points,
with two sets of $\sigma-$dependent constraints using Taylor expansion.
We found it convenient to separate sums with even and odd
powers of $\sigma$, because $\Omega$ symmetric and antisymmetric functions are independent.
Note that these constraints are infinite sums,
bilinear in coordinate, with one $\tau$ derivative and
an arbitrary degree of $\sigma-$derivatives.
The main formulas were derived in Appendixes \ref{sec:iform}-\ref{sec:hcompact}.

This stage was a good point for transition from Lagrangian to Hamiltonian method.
We expressed the $\tau-$derivative
of a coordinates (${\dot{x}}^\mu$),
in terms of momenta $\pi_\mu$ and the $\sigma-$derivative ($x^{\prime\mu}$),
and obtained constraints in the Hamiltonian form.
Then, we were in position to check the validity of our procedure.
Because the $\Omega$ even and odd
parts of the constraints are independent,
it was useful to consider their difference as the single constraint.
The Poisson brackets between the Hamiltonian and this constraint
are just the $\sigma-$derivative of the constraint.
It means that it weakly commutes with the Hamiltonian.
First, this proved that the expression obtained
from the boundary conditions with the help of the Lagrangian consistency procedure is
really Hamiltonian constraint. Second, we concluded that there were
no more constraints, and the consistency procedure was completed.
So, we showed the equivalence with the standard Dirac consistency procedure by
rewriting the constraints in the canonical form.

The Poisson brackets between constraints in the leading order are closed on the metric tensor
times the $\sigma$ derivative of the $\delta-$function.
The metric tensor is regular (det $G_{\mu\nu}\neq 0$), and,
consequently, all constraints
except the zero modes \cite{BS} are of the second class.
There are two possibilities to deal with second class constraints.
The usual approach is to find the Dirac bracket,
but, in our particular case, it is simpler to solve them explicitly.
As a consequence of $\sigma-$derivative of $\delta-$function,
zero modes of constraints are of the first class.
They are generators of global symmetry,
which we used to gauge away center of mass of the coordinate.

The simple solution (\ref{eq:solution}) of the original
closed-string coordinates $x^\mu$,
in terms of effective open-string
coordinates $q^\mu$ and momenta $p_\nu$,
defines the noncommutative
product and its properties. The noncommutativity parameter
$\theta^{\mu\nu}$, (\ref{eq:teta}), formally has the same form as
in the flat background, but with Kalb-Ramond field,
linear in coordinate, (\ref{eq:gb}).
Taking into account the fact that, at the
string boundaries, $\bar{q}(0)=0$ and $\bar{q}(\pi)=0$ and gauging
away canter of mass coordinate, we can rewrite explicitly
the noncommutative relations in terms of the original variables
\begin{eqnarray}\label{eq:stc}
{}^{\star}\{X^\mu(0),X^\nu(0)\}&=&-\theta^{\mu\nu}[X(0)]=f^{\mu\nu}_{\ \ \rho} X^\rho(0),
\nonumber\\
{}^{\star}\{X^\mu(\pi),X^\nu(\pi)\}&=&\theta^{\mu\nu}[X(\pi)]=-f^{\mu\nu}_{\ \ \rho} X^\rho(\pi),  \
\end{eqnarray}
where $f^{\mu\nu}_{\ \ \rho}=\frac{2}{3\kappa}B^{\mu\nu}_{\ \ \rho}$.

In Refs. \cite{CSS,HKK},
the noncommutative product is defined
only on the world-sheet boundary
using path integral method.
In fact, the explicit expression of star product has been extracted
from the correlation functions computed on the disk.

On the other hand,
we used canonical approach and explicitly solved
boundary conditions. We want to stress that
only space-time coordinates of the string endpoints are
noncommutative. For any points of the string interior,
the commutation relations are standard.
This is obvious in a decoupling limit
($\alpha^\prime\rightarrow\sqrt{\varepsilon}\alpha^\prime$,
$G_{\mu\nu}\rightarrow\varepsilon G_{\mu\nu}$, $\varepsilon\rightarrow{0}$),
when all degrees of freedom in the string interior can be gauged away \cite{CF}.
We could also expect such a result,
without decoupling limit,
but for the constant $B_{\mu\nu}$ field,
because it does not appear in equations of motion
and do not affect the string
interior \cite{ACL,SN}. In our case, this is nontrivial result, because
the coordinate-dependent $B_{\mu\nu}$ field contributes to equations of motion
and affects string interior.
Furthermore, even at the first glance,
one sees that the constraints can not be imposed only on the string endpoints.
In fact, $\gamma^\mu_{0}$, which are defined only on the boundary,
are just prime constraints.
In order to obtain all the constraints,
one must apply the Dirac consistency procedure,
which leads to the full set of constraints $\Gamma^\mu(\sigma)$,
which is nontrivial at the string interior.
So, even in the case when the term with the Kalb-Ramond field in the action is not topological,
it is possible to restrict noncommutativity only to the world-sheet boundary.

Let us now discuss the relation between our case of branes in weakly curved background with
branes on group manifold \cite{Sch,VS,ARS}.
Note that the strings moving on group manifold
are described by Wess-Zumino-Novikov-Witten (WZNW) model.
Owing to conformal symmetries of the  WZNW model, the space-time Eqs.
(\ref{eq:beta})-(\ref{eq:beta1}) are automatically satisfied.

Strings moving on 3-sphere $S^{3}$ of radius $R$ are described by WZNW model,
with group $SU(2)$ at level $n$.
As a consequence of Dirac quantization condition,
it follows that radius of 3-sphere is quantized,
$R^{2}=\alpha^\prime n$, where the integer $n$ is also the level of the corresponding current algebra.
In the limit of large $n$, the group manifold becomes more and more flat,
and 3-sphere approaches the flat 3-space.
So, in the language of the group manifold,
the large level $n$ corresponds to the weakly curved
background of the present paper. In that sense, our result (\ref{eq:stc}) corresponds
to Eq. (\ref{eq:poib}) of Ref.\cite{VS}, and
the structure constants $f^{\mu\nu}_{\ \ \rho}$ are proportional
to the field strength of the Kalb-Ramond field $B^{\mu\nu}_\rho$.

As mentioned in Ref.\cite{VS},
these relations have been obtained as an extension of the flat-background expressions
and "naively applied" to curved background.
In the present paper, we derive the Eq. (\ref{eq:poib}) of Ref. \cite{VS}
and prove that it is correct.
Our derivation is not restricted to the case of $SU(2)$ group.

After quantization, using Weyl normal ordering prescription, we showed that the product
of operators, defined on Dp-brane world-volume, is isomorphic to Kontsevich product of ordinary functions.
In the case of weakly curved background,
Kontsevich product turns to associative one,
because nonassociative term is infinitesimal of the second order.

Let us discuss one additional possibility in our approach.
Instead of using the composed variables $X^\mu$ with noncommutativity relation
(\ref{eq:com}) and the Weyl normal ordering prescription (\ref{eq:weyl}),
we can treat $q^\mu$ and $p_\mu$ as fundamental variables.
Then, we can define normal ordering :: for operators ${\hat{q}}^\mu$ and ${\hat{p}}_\mu$ and,
to any function $f(x)$,
according to (\ref{eq:solution}), assign the operator
\begin{equation}\label{eq:norm}
f(x)\rightarrow
:\hat{f}\{{\hat{q}}^\mu-\int_{0}^{\sigma}d\sigma_{0}[\theta^{\mu\nu}(\hat{q}){\hat{p}}_\nu
+\frac{1}{2}\theta^{\prime\mu\nu}(\hat{q}){\hat{P}}_\nu]\}:.
\end{equation}

Consequently, we can introduce the new star product,
specifying new normal ordering and using
Eq. (\ref{eq:norm}) and the commutation relation (\ref{eq:poisson}).
The new star product is defined along the whole string
and not only on the string endpoints.
We find this approach more fundamental but, in the particular case
on the world-sheet boundary,
it produces the same Kontsevich star product.
We will discuss this new definition of star product elsewhere.

In the present paper,
in order to simplify calculations,
we neglected the constant part $b_{\mu \nu}$  of the linear Kalb-Ramond
field $B_{\mu \nu} = b_{\mu \nu} + {1  \over
3}B_{\mu\nu\rho}x^\rho$, introduced in Eq. (\ref{eq:gb}).
In that case, the
new, momentum-dependent term of Ref.\cite{HY} goes to zero.
This resolves the second ambiguity from the Introduction, that
the result of Refs.\cite{CSS,HKK} is valid for $b_{\mu \nu}= 0$.

From where does
the momentum-dependent term appear?
It can come from the Poisson bracket between the momentum-dependent
terms of the relation (\ref{eq:solution}),
the basic expression of the initial coordinate $x^\mu$ in terms of
effective canonical variables $q^\mu$ and $p_\mu$.
In the present paper (for $b_{\mu
\nu}= 0$), this part is infinitesimal of the second order, so we neglect it.
In the case $b_{\mu \nu}\neq 0$, this part
produces a nontrivial momentum-dependent result,
because
the expression $\theta^{\mu\nu}(q)$
acquires the constant finite term $\theta^{\mu\nu}_{0}\neq 0$.

In our next paper, Ref.\cite{DS},
we apply the same canonical
method for the case $b_{\mu \nu} \neq 0$ and obtain momentum-dependent
noncommutativity parameter. Beside the standard
expression of Refs.\cite{CSS,HKK}, it contains the term of
Ref.\cite{HY} and some other momenta dependent terms. This result
will resolve all the ambiguities mentioned in Introduction and it
represents a complete expression of the noncommutativity parameter of the weakly
curved background.

\appendix
\section{$2\pi-$ periodic functions}\label{sec:deltateta}
\cleq
In this Appendix, we will introduce the Fourier expansion of the ordinary,
symmetric, and antisymmetric delta and step functions.
In addition, we define $I_{k}$ functions as $k$ integrals of the symmetric
$\theta$ functions and investigate their properties.
\subsection{Step and delta functions}
\cleq
The Fourier series of the $2\pi$-periodic $\delta$-function
and the $\theta$ step function have the forms
\begin{equation}\label{eq:delta}
\delta(\sigma)=\frac{1}{2\pi}+\frac{1}{\pi}\sum_{n\geq 1}\cos{n\sigma},\quad\quad(\sigma\in[0,2\pi])
\end{equation}
\begin{equation}\label{eq:step}
\theta(\sigma)=\frac{1}{2\pi}\Big{(}\sigma+2\sum_{n\geq 1}
\frac{1}{n}\sin{n\sigma}\Big{)},
\end{equation}
where,
by definition,
$\theta(\sigma)=\int_{0}^{\sigma}d\sigma_{0}\delta(\sigma_{0})$.
Let us define the delta and step functions, symmetric and antisymmetric,
under $\sigma$-parity:
\begin{eqnarray}\label{eq:dsa}
\delta_{S}(\sigma,\bar{\sigma})=
\frac{1}{2}[\delta(\sigma-\bar{\sigma})+\delta(\sigma+\bar{\sigma})],&&
\delta_{A}(\sigma,\bar{\sigma})=
\frac{1}{2}[\delta(\sigma-\bar{\sigma})-\delta(\sigma+\bar{\sigma})],
\nonumber\\
\theta_{S}(\sigma,\bar{\sigma})=
\frac{1}{2}[\theta(\sigma-\bar{\sigma})+\theta(\sigma+\bar{\sigma})],&&
\theta_{A}(\sigma,\bar{\sigma})=
\frac{1}{2}[\theta(\sigma-\bar{\sigma})-\theta(\sigma+\bar{\sigma})].
\end{eqnarray}
Using (\ref{eq:delta}) and (\ref{eq:step}),
we can rewrite them in the form
\begin{eqnarray}
\delta_{S}(\sigma,\bar{\sigma})=
\frac{1}{2\pi}\Big{[}1+2\sum_{n\geq{1}}\cos{n\sigma}\cos{n\bar{\sigma}}\Big{]},&&
\delta_{A}(\sigma,\bar{\sigma})=
\frac{1}{\pi}\sum_{n\geq{1}}\sin{n\sigma}\sin{n\bar{\sigma}},
\nonumber\\
\theta_{S}(\sigma,\bar{\sigma})=
\frac{1}{2\pi}\Big{[}\sigma+2\sum_{n\geq{1}}\frac{1}{n}\sin{n\sigma}\cos{n\bar{\sigma}}\Big{]},&&
\theta_{A}(\sigma,\bar{\sigma})
=-\frac{1}{2\pi}\Big{[}\bar{\sigma}+2\sum_{n\geq{1}}\frac{1}{n}\cos{n\sigma}\sin{n\bar{\sigma}}\Big{]}.
\nonumber\\
\end{eqnarray}
These functions satisfy the following properties:
\begin{eqnarray}
\delta_{S}(\sigma,\bar{\sigma})=\delta_{S}(\bar{\sigma},\sigma),&&
\delta_{A}(\sigma,\bar{\sigma})=\delta_{A}(\bar{\sigma},\sigma),
\nonumber\\
\delta_{S}(\sigma,-\bar{\sigma})=\delta_{S}(\sigma,\bar{\sigma}),&&
\delta_{A}(\sigma,-\bar{\sigma})=-\delta_{A}(\sigma,\bar{\sigma}),
\nonumber\\
\theta_{S}(\bar{\sigma},\sigma)=-\theta_{A}(\sigma,\bar{\sigma}),&&
\partial_{\sigma}\theta_{S}(\sigma,\bar{\sigma})=
\delta_{S}(\sigma,\bar{\sigma}),
\nonumber\\
\partial_{\sigma}\theta_{A}(\sigma,\bar{\sigma})=
\delta_{A}(\sigma,\bar{\sigma}),&&
\partial_{\bar{\sigma}}\theta_{S}(\sigma,\bar{\sigma})=
-\delta_{A}(\sigma,\bar{\sigma}).
\end{eqnarray}
We will use the relations
\begin{eqnarray}
\int_{0}^{\sigma}d\sigma_{1}f(\sigma_{1})\delta(\sigma_{1}-\bar{\sigma})&=&
f(\bar{\sigma})[\theta(\sigma-\bar{\sigma})+\theta(\bar{\sigma})],
\nonumber\\
\int_{0}^{\sigma}d\sigma_{1}f_{S}(\sigma_{1})\delta_{S}(\sigma_{1},\bar{\sigma})&=&
f_{S}(\bar{\sigma})\theta_{S}(\sigma,\bar{\sigma}),
\nonumber\\
\int_{0}^{\sigma}d\sigma_{1}f_{A}(\sigma_{1})\delta_{A}(\sigma_{1},\bar{\sigma})&=&
f_{A}(\bar{\sigma})\theta_{S}(\sigma,\bar{\sigma}),
\end{eqnarray}
where $f_{S}$ and $f_{A}$ are symmetric and antisymmetric functions under
$\sigma$-parity,
$f_{S}(-\sigma)=f_{S}(\sigma)$ and
$f_{A}(-\sigma)=-f_{A}(\sigma)$.
Using the fact that
\begin{equation}
\theta(\sigma)= \left\{ \begin{array}{rcl}
0 && \sigma=0\\
1/2 && 0 < \sigma < 2\pi\\
1 && \sigma=2\pi
\end{array}\right.,
\qquad\qquad(\sigma\in[0,2\pi])
\end{equation}
we obtain
\begin{equation}\label{eq:thetas}
\theta_{S}(\sigma,\bar{\sigma})= \left\{ \begin{array}{rcl}
0 && \sigma=\bar{\sigma}=0\\
1/2 && \sigma=\bar{\sigma}=\pi\\
1/4 && \sigma=\bar{\sigma}\neq 0,\pi\\
1/2 && \sigma>\bar{\sigma}\\
0 && \sigma<\bar{\sigma}\\
\end{array}\right.,\qquad\qquad(\sigma,\bar{\sigma}\in [0,\pi])
\end{equation}
which will be useful in derivation the properties of the $I_{k}$ functions.
\subsection{Integrals of the symmetric $\theta$ functions}
We define two variable functions $I_{k}(\sigma,\bar{\sigma})$,
($\sigma,\bar{\sigma}\in[0,\pi]$) as multiple integrals of a symmetric step function
\begin{eqnarray}\label{eq:ikdef}
I_{0}(\sigma,\bar{\sigma})&=&\theta_{S}(\sigma,\bar{\sigma}),
\nonumber\\
I_{k}(\sigma,\bar{\sigma})&=&
\int_{0}^{\sigma}d{\sigma_{1}}^{2}
\int_{0}^{\sigma_{1}}d{\sigma_{2}}^{2}\cdots
\int_{0}^{\sigma_{k-1}}d\sigma_{k}^{2}
\theta_{S}(\sigma_{k},\bar{\sigma}).\qquad\qquad(k\geq 1)
\end{eqnarray}
They have the following properties:
\begin{eqnarray}\label{eq:dik}
\partial_\sigma I_{k}(\sigma,\bar{\sigma})&=&2\sigma I_{k-1}(\sigma,\bar{\sigma}),
\nonumber\\
\partial_{\bar{\sigma}} I_{k}(\sigma,\bar{\sigma})&=&-2\bar{\sigma} I_{k-1}(\sigma,\bar{\sigma}).\qquad(k\geq 1)
\end{eqnarray}
Using (\ref{eq:thetas}) and the mathematical induction,
it can be shown that
\begin{eqnarray}\label{eq:ik}
I_{k}(\sigma,\bar{\sigma})&=& \left\{ \begin{array}{rcl}
\frac{1}{2k!}(\sigma^{2}-{\bar{\sigma}}^{2})^{k} && \sigma>{\bar{\sigma}}\\
0 && \sigma\leq{\bar{\sigma}}\\
\end{array}\right. .
\qquad\quad (k \geq 1)
\end{eqnarray}
In the derivation of the summation formula in Appendix \ref{sec:summ},
we will need an expression
for the $k-$th derivative ($k\leq{n}$)
of the $I_{n}$ function over second variable.
The expression $\partial^{k}_{\bar{\sigma}}I_{n}(\sigma,\bar{\sigma})$ is a polynomial of the $(2n-k)$-th order
in $\bar{\sigma}$, so it can be written in a form
\begin{equation}\label{eq:dkIn}
\partial^{k}_{\bar{\sigma}}I_{n}(\sigma,\bar{\sigma})=
\sum_{q=0}^{[k/2]}a^{k}_{q}{\bar{\sigma}}^{k-2q}I_{n-k+q}(\sigma,\bar{\sigma}).
\end{equation}
Using the mathematical induction,
we obtain the recursion relation for
coefficients $a^{k}_{q}$, with the solution
\begin{eqnarray}\label{eq:a}
a^{k}_{0}&=&(-2)^{k},\quad k \geq 0
\nonumber\\
a^{k}_{q}&=&(-2)^{k-q}{k \choose {2q}}(2q-1)!!\qquad\qquad (k \geq 2q).
\end{eqnarray}
\section{ Induced brackets in the reduced phase space}\label{sec:starb}
\cleq
The solution of the constraints  $\Gamma_\mu(\sigma)=0$ and
$\bar\Gamma_\mu(\sigma)=0$ reduces the phase space,
leaving only half of the degrees of freedom.
Let us clarify the relation between the
brackets associated with the initial and the reduced  phase spaces.
We will distinguish two nontrivial steps.
In Appendix \ref{sec:b2},
we will take into account the symmetries of the basic
canonical variables under $\sigma$-parity $\Omega$,
and, in Appendix \ref{sec:b1},
we will impose second class constraints $\Gamma_\mu$.
\subsection{The phase space reduced by $\Omega$-even and $\Omega$-add projections}
\label{sec:b2}

First,
we need the expression for brackets between
basic canonical variables $q^\mu$ and $p_\mu$
(${\bar{q}}^\mu$ and ${\bar{p}}_\mu$) in the interval
$[0,\pi]$. Note that they are not closed on standard $\delta$
function, because they are not arbitrary functions on that
interval, but they contain only even (odd) powers of $\sigma$.
The easiest way to impose this restriction is just an extension to the
domain $[-\pi,\pi]$, when they become symmetric (antisymmetric)
functions under $\sigma \to - \sigma$. Then, we have
\begin{equation}
\{q^\mu(\sigma),p_\nu(\bar{\sigma})\}=
\delta^{\mu}_{\nu} \delta_{S}(\sigma,\bar{\sigma}),\qquad
\{{\bar{q}}^\mu(\sigma),{\bar{p}}_\nu(\bar{\sigma})\}=
\delta^{\mu}_{\nu} \delta_{A}(\sigma,\bar{\sigma}),
\end{equation}
with $\sigma,
\bar{\sigma} \in[-\pi,\pi]$
where, by definition,
\begin{equation}
\int_{-\pi}^{\pi} d \bar\sigma q^\mu(\bar\sigma)
\delta_{S}(\bar{\sigma},\sigma)= q^\mu(\sigma),\qquad
\int_{-\pi}^{\pi} d \bar\sigma {\bar{q}}^\mu(\bar\sigma)
\delta_{S}(\bar{\sigma},\sigma)= {\bar{q}}^\mu(\sigma).
\end{equation}

Separating integration domain in two parts, from $-\pi$ to $0$ and
from $0$ to $\pi$, and changing the integration variable in the first
part $\bar{\sigma} \to - \bar{\sigma}$, we obtain
\begin{equation}
2\int_0^{\pi} d \bar\sigma q^\mu(\bar\sigma)
\delta_{S}(\bar{\sigma},\sigma)
= q^\mu(\sigma),\qquad
2\int_0^{\pi} d \bar\sigma {\bar{q}}^\mu(\bar\sigma)
\delta_{S}(\bar{\sigma},\sigma)
= {\bar{q}}^\mu(\sigma).
\end{equation}

So, the unit functions on the interval $[0,\pi]$ for
functions with only an even or odd power in $\sigma$ are
$2 \delta_{S}(\bar{\sigma},\sigma)$
and
$2 \delta_{A}(\bar{\sigma},\sigma)$, respectively.
Therefore, the brackets
which we are looking for
have a form
\begin{equation}\label{eq:brac}
\{q^\mu(\sigma),p_\nu(\bar{\sigma})\}= 2
\delta^{\mu}_{\nu} \delta_{S}(\sigma,\bar{\sigma}),\quad
\{{\bar{q}}^\mu(\sigma),{\bar{p}}_\nu(\bar{\sigma})\}= 2
\delta^{\mu}_{\nu} \delta_{A}(\sigma,\bar{\sigma}),
\quad \sigma,
\bar{\sigma} \in[0,\pi].
\end{equation}

In the initial phase space,
with the canonical variables
$x^\mu(\sigma)$, $\pi_\mu(\sigma)$,and $\sigma\in[0,\pi]$,
the standard Poisson bracket (\ref{eq:poib}) is valid. Applying
relations (\ref{eq:brac}), we obtain Poisson brackets (\ref{eq:poisson}).
\subsection
{The phase space reduced by the constraint
$\Gamma_\mu=0$}\label{sec:b1}

On the solution of boundary conditions, we obtain reduced phase space with
$2\pi$-periodic canonical variables $q^\mu(\sigma)$ and
$p_\mu(\sigma)$, defined in (\ref{eq:qqbar}) and (\ref{eq:qp}).
For arbitrary functions $F(x,\pi)$ and $G(x,\pi)$, defined
on the initial phase space, we introduce their restrictions on the
reduced phase space, as a value on the solution of the
boundary conditions
\begin{equation}
f(q,p)=F(x,\pi)\Big{|}_{\Gamma_\mu=0}, \qquad
g(q,p)=G(x,\pi)\Big{|}_{\Gamma_\mu=0}.
\end{equation}

As was shown in the Sec. 2.3.2
of the Ref. \cite{HT},
the Poisson brackets in the effective phase
space are, in fact, the Dirac brackets in the initial phase space
associated with the second class constraints $\Gamma_\mu=0$,
\begin{equation}\label{eq:starb}
{}^{\star}\{f,g\}=\{F,G\}_{Dirac}\Big{|}_{\Gamma_\mu=0}.
\end{equation}
To distinguish the new brackets from that of the initial phase
space, we denoted them by star.
Applying the first relation (\ref{eq:brac}) to the star brackets (\ref{eq:starb}),
we obtain (\ref{eq:stbrac}).

Finally, we should check the $2\pi$-periodicity conditions (\ref{eq:dvapi}).
For the $\sigma$-symmetric functions ($q^\mu(\sigma)$ and
$p_\mu(\sigma)$ and their algebraic combinations) they are
automatically satisfied. The $\sigma$-antisymmetric functions
(the $\sigma$-derivative and $\sigma$-integral of the symmetric
functions) must vanish
both at $\sigma=0$ and $\sigma=\pi$
because of the antisymmetry and
$2\pi$-periodicity, respectively.
\section{Integral form of $({\Gamma}^{Q})^{\alpha\beta}$ and $({\bar{\Gamma}}^{Q})^{\alpha\beta}$}\label{sec:iform}
\cleq
\subsection{The case $\sigma=0$}\label{sec:ioform}
We will show that $({\Gamma}^{Q})^{\alpha\beta}$,
defined in (\ref{eq:gsum}), is equal to
\begin{equation}\label{eq:gqint}
(\Gamma^{Q})_{k}^{\ \alpha\beta}(\sigma)=
\frac{(-1)^{k}\sigma}{2(k+1)!}
\int_{0}^{\sigma}d\sigma_{1}^{2}
\int_{0}^{\sigma_{1}}d\sigma_{2}^{2}\cdots
\int_{0}^{\sigma_{k-1}}d\sigma_{k}^{2}
(Q_{A})_{k}^{\alpha\beta}(\sigma_{k}).
\end{equation}
In order to prove the above relation,
it is useful to define auxiliary variable
\begin{eqnarray}
(\gamma^{Q})^{\alpha\beta}_{kq}(\sigma)=
\sum_{n=k+1}^{\infty}\frac{\sigma^{2n-2q-1}}{(2n-2q-1)!}
\frac{(n-q-1)!}{(n-k-1)!}
Q_{k}^{\alpha\beta(2n-2k-1)}\Big{|}_{\sigma=0},\quad (q=0,1,\cdots ,k)
\end{eqnarray}
and rewrite $(\Gamma^{Q})_{k}^{\alpha\beta}$ as
\begin{equation}
(\Gamma^{Q})_{k}^{\alpha\beta}=(-4)^{k}\frac{\sigma}{2(k+1)!}
(\gamma^{Q})^{\alpha\beta}_{k0}(\sigma).
\end{equation}
Observing that $(\gamma^{Q})^{\alpha\beta}_{kq}$ satisfies
\begin{equation}
(\gamma^{Q})^{\prime\alpha\beta}_{kq}(\sigma)=
\frac{\sigma}{2}(\gamma^{Q})^{\alpha\beta}_{k{q+1}}(\sigma)
\quad\Rightarrow
(\gamma^{Q})^{\alpha\beta}_{kq}(\sigma)=
\frac{1}{4}\int_{0}^{\sigma}d\sigma_{1}^{2}
(\gamma^{Q})^{\alpha\beta}_{k{q+1}}(\sigma_{1})
\end{equation}
and using the fact that
\begin{eqnarray}\label{eq:q}
(\gamma^{Q})^{\alpha\beta}_{kk}&=&
(Q_{A})_{k}^{\alpha\beta}(\sigma)\equiv
\sum_{n=0}^{\infty}\frac{{\sigma}^{2n+1}}{(2n+1)!}
Q_{k}^{(2n+1)\alpha\beta}\Big{|}_{0}
=[{\dot{x}}^{(k)\alpha}x^{(k+1)\beta}]_{A},
\end{eqnarray}
is the antisymmetric part of $Q^{\alpha\beta}_{k}$, we obtain (\ref{eq:gqint}).
\subsection{The case $\sigma=\pi$}\label{sec:ipform}
A similar integral form can be obtained for $({\bar{\Gamma}}^{Q})^{\alpha\beta}$,
defined in (\ref{eq:gqnad}).
As before, it is useful to define auxiliary variable as
\begin{eqnarray}
({\bar{\gamma}}^{Q})^{\alpha\beta}_{kq}(\sigma)=
\sum_{n=k+1}^{\infty}\frac{(\sigma-\pi)^{2n-2q-1}}{(2n-2q-1)!}
\frac{(n-q-1)!}{(n-k-1)!}
Q_{k}^{\alpha\beta(2n-2k-1)}\Big{|}_{\sigma=\pi},\; (q=0,1,\cdots ,k)
\end{eqnarray}
and rewrite $({\bar{\Gamma}}^{Q})_{k}^{\alpha\beta}$ as
\begin{equation}
({\bar{\Gamma}}^{Q})_{k}^{\alpha\beta}=(-4)^{k}\frac{\sigma-\pi}{2(k+1)!}
({\bar{\gamma}}^{Q})^{\alpha\beta}_{k0}(\sigma).
\end{equation}
Observing that $({\bar{\gamma}}^{Q})^{\alpha\beta}_{kq}$ satisfies
\begin{equation}
({\bar{\gamma}}^{Q})^{\prime\alpha\beta}_{kq}(\sigma)=
\frac{\sigma-\pi}{2}({\bar{\gamma}}^{Q})^{\alpha\beta}_{k{q+1}}(\sigma)
\Rightarrow
({\bar{\gamma}}^{Q})^{\alpha\beta}_{kq}(\sigma)=
\int_{\sigma}^{\pi}d\sigma_{1}
\frac{\pi-\sigma_{1}}{2}
({\bar{\gamma}}^{Q})^{\alpha\beta}_{k{q+1}}(\sigma_{1}),
\end{equation}
we obtain
\begin{equation}\label{eq:bgq}
({\bar{\Gamma}}^{Q})_{k}^{\ \alpha\beta}(\sigma)=
\frac{\sigma-\pi}{2(k+1)!}
\int_{\sigma}^{\pi}d(\sigma_{1}-\pi)^{2}
\int_{\sigma_{1}}^{\pi}d(\sigma_{2}-\pi)^{2}\cdots
\int_{\sigma_{k-1}}^{\pi}d(\sigma_{k}-\pi)^{2}
({\bar{Q}}_{A})_{k}^{\alpha\beta}(\sigma_{k}),
\end{equation}
with
\begin{equation}
({\bar{\gamma}}^{Q})^{\alpha\beta}_{kk}(\sigma)=({\bar{Q}}_{A})^{\alpha\beta}_{k}(\sigma)
\equiv\sum_{n=0}^{\infty}\frac{(\sigma-\pi)^{2n+1}}{(2n+1)!}
{Q_{k}}^{(2n+1)\alpha\beta}\Big{|}_{\sigma=\pi}.
\end{equation}
\section{Summation formula}\label{sec:summ}
\cleq
In this Appendix, we will derive the relation
\begin{eqnarray}\label{eq:sf}
S^\rho(x|\sigma,\bar{\sigma})&\equiv &\sum_{k=0}^{\infty}
\frac{1}{(k+1)!}
\frac{\partial^{k}}{\partial{\bar{\sigma}}^{k}}
\Big{[}\bar{\sigma} x^{(k+1)\rho}(\bar{\sigma})I_{k}({\sigma},\bar{\sigma})\Big{]}
\nonumber\\
&=&\frac{1}{2}\theta_{S}(\sigma,\bar{\sigma})
[x^\rho(\sigma)+x^\rho(-\sigma)-2x^\rho(-\bar{\sigma})],
\end{eqnarray}
which we will use in the Appendix \ref{sec:hcompact}.

Using the Leibniz rule and Eq. (\ref{eq:dkIn}), the sum in the above
expression can be rewritten as an expansion in $I_{k}$ functions,
\begin{equation}\label{eq:es}
S^\rho(x|\sigma,\bar{\sigma})=\sum_{l=0}^{\infty}C^\rho_{l}(\bar{\sigma})I_{l}(\sigma,\bar{\sigma}),
\end{equation}
with the coefficients
\begin{equation}\label{eq:cfro}
C^{\rho}_{l}(\bar{\sigma})=
\sum_{q=0}^{l}\sum_{k=2l-q}^{\infty}
\frac{1}{(k+1)!}{k \choose{q}}\Big{[}\bar{\sigma}x^{\rho(k+1)}(\bar{\sigma})\Big{]}^{(q)}
a^{k-q}_{l-q}{\bar{\sigma}}^{k-2l+q}.
\end{equation}
For $l=0$, we obtain
\begin{equation}\label{eq:c0}
C_{0}^{\rho}=\frac{1}{2}[x^\rho(\bar{\sigma})-x^\rho(-\bar{\sigma})].
\end{equation}
With the help of (\ref{eq:a}),
we can rewrite the coefficients (\ref{eq:cfro}) in the form
\begin{equation}\label{eq:cfro1}
C^\rho_{l}(\bar{\sigma})=\sum_{m=0}^{\infty}K_{lm}\frac{(-2\bar{\sigma})^{m}}{m!}x^{(m+2l)\rho},\quad(l\geq{1})
\end{equation}
where
\begin{eqnarray}\label{eq:ka}
K_{l0}&=&\frac{(-1)^{l}}{l!}(2l+1)R_{2l+1,l},
\nonumber\\
K_{lm}&=&\frac{(-1)^{l}}{l!}\frac{(2l+m)}{2}R_{2l+m,l},\qquad(m\geq 1)
\end{eqnarray}
and $R_{m,n}$ are defined in Appendix \ref{sec:rcof}.
With the help of (\ref{eq:rel}), we can rewrite (\ref{eq:cfro1}) as
\begin{eqnarray}\label{eq:coef}
C^\rho_{l}(\bar{\sigma})&=&\frac{1}{2}\sum_{m=0}^{\infty}
\frac{(-2\bar{\sigma})^{m}}{m!}
\frac{(m+l-1)!}{(m+2l-1)!}x^{(m+2l)\rho}(\bar{\sigma}).\quad\quad (l\geq{1})
\end{eqnarray}
Let us now define the auxiliary function of two variables
\begin{equation}
{\bar{C}}^{\rho}_{l}(\eta,\bar{\sigma})=\frac{1}{2}\sum_{m=0}^{\infty}
\frac{(-2\eta)^{m}}{m!}
\frac{(m+l-1)!}{(m+2l-1)!}x^{(m+2l)\rho}(\bar{\sigma}).\quad\quad (l\geq 1)
\end{equation}
Obviously, $C^\rho_{l}(\bar{\sigma})={\bar{C}}^\rho_{l}(\bar{\sigma},\bar{\sigma})$.
Let us define
\begin{eqnarray}
{\bar{C}}^{k\rho}_{l}(\eta_{k},\bar{\sigma})&=&
\int_{0}^{\eta_{k}}d\eta_{k-1} {\bar{C}}^{k-1\rho}_{l}(\eta_{k-1},\bar{\sigma}),
\quad (1\leq k \leq l-1)
\nonumber\\
{\bar{C}}^{0\rho}_{l}(\eta_{k},\bar{\sigma})&=&{\bar{C}}^{\rho}_{l}(\eta_{k},\bar{\sigma}).
\end{eqnarray}

We can show that ${\bar{C}}^{l-1\rho}_{l}$ is equal to
\begin{equation}
{\bar{C}}^{l-1\rho}_{l}(\eta,\bar{\sigma})=
-\frac{1}{(4\eta)^{l}}
\Big{[}x^{\prime\rho}(\bar{\sigma}-2\eta)
-\sum_{n=0}^{2l-2}
\frac{(-2\eta)^{n}}{n!}x^{(n+1)\rho}(\bar{\sigma})\Big{]},
\end{equation}
and, using
\begin{equation}
{\bar{C}}^{\rho}_{l}(\eta,\bar{\sigma})=
\partial_{\eta}^{\ l-1}{\bar{C}}^{\ l-1\rho}_{l}(\eta,\bar{\sigma}),\\
\end{equation}
we obtain
\begin{eqnarray}
{\bar{C}}^{\rho}_{l}(\eta,\bar{\sigma})&=&
\frac{(-1)^{l}}{2}\sum_{n=0}^{l-1}\frac{(2l-n-2)!}{n!(l-n-1)!}
(2\eta)^{-(2l-n-1)}
[x^{(n+1)\rho}(\bar{\sigma}-2\eta)-(-1)^{n}x^{(n+1)\rho}(\bar{\sigma})],
\nonumber\\
\end{eqnarray}
and, finally,
\begin{equation}\label{eq:cro}
{C}^{\rho}_{l}(\bar{\sigma})=
\frac{(-1)^{l}}{2}\sum_{n=0}^{l-1}\frac{(2l-n-2)!}{n!(l-n-1)!}
(2\bar{\sigma})^{-(2l-n-1)}
[x^{(n+1)\rho}(-\bar{\sigma})-(-1)^{n}x^{(n+1)\rho}(\bar{\sigma})].
\end{equation}
Substituting (\ref{eq:cro}) and (\ref{eq:ik}) into
expression (\ref{eq:es}) without the first term
\begin{equation}\label{eq:s1}
S^{\rho}_{1}(x|\sigma,\bar{\sigma})=\sum_{l=1}^{\infty}C^\rho_{l}(\bar{\sigma})I_{l}(\sigma,\bar{\sigma}),
\end{equation}
after straightforward calculation for $\sigma>\bar{\sigma}$, we obtain
\begin{eqnarray}\label{eq:s1ro}
S^{\rho}_{1}(x|\sigma,\bar{\sigma})&=&
\frac{1}{4}\sum_{n=0}^{\infty}\frac{(-2\bar{\sigma})^{n+1}}{n!}
\Big{[}(-1)^{n+1}x^{(n+1)\rho}(-\bar{\sigma})+x^{(n+1)\rho}(\bar{\sigma})\Big{]}
{\tilde{S}}_{n}(y),
\nonumber\\
y&=&\frac{1}{4}\Big{[}1-\Big{(}\frac{\sigma}{\bar{\sigma}}\Big{)}^2\Big{]},
\end{eqnarray}
where ${\tilde{S}}_{n}(y)$ is defined in (\ref{eq:stilda}).
Substituting (\ref{eq:stn}) into (\ref{eq:s1ro}), we obtain
\begin{eqnarray}
S^{\rho}_{1}(x|\sigma,\bar{\sigma})&=&
\frac{1}{4}\sum_{n=0}^{\infty}\frac{1}{(n+1)!}
(\sigma-\bar{\sigma})^{n+1}
[x^{(n+1)\rho}(\bar{\sigma})+(-1)^{n+1}x^{(n+1)\rho}(-\bar{\sigma})]
\nonumber\\
&=&\frac{1}{4}[x^{\rho}({\sigma})+x^{\rho}(-{\sigma})
-x^{\rho}(\bar{\sigma})-x^{\rho}(-\bar{\sigma})]\quad (\sigma>\bar{\sigma}).
\end{eqnarray}
Using the properties of $\theta$ functions and Eq. (\ref{eq:c0}),
we obtain (\ref{eq:sf}).
\subsection{ Coefficients $R_{m,n}$}\label{sec:rcof}
In this Appendix, we will prove the relation
\begin{eqnarray}\label{eq:rel}
R_{m,n}&\equiv & \sum_{k=0}^{n}{n \choose k}\frac{(-1)^{k}}{m-k}
=(-1)^{n}\frac{n!(m-n-1)!}{m!}\quad (m>n),
\end{eqnarray}
used in (\ref{eq:ka}). Let us introduce auxiliary function
\begin{eqnarray}\label{eq:fmn}
f_{mn}(\alpha)&\equiv & \sum_{k=0}^{n}{n \choose k}\frac{(-1)^{k}}{m-k}\alpha^{m-k},\quad (m>n)
\end{eqnarray}
which has properties
\begin{eqnarray}\label{eq:border}
f_{mn}(0)=0, && f_{mn}(1)=R_{m,n}.
\end{eqnarray}
Differentiating $f_{mn}$ over $\alpha$, we obtain
\begin{equation}
f^{\prime}_{mn}(\alpha)=(-1)^{n}\alpha^{m-n-1}(1-\alpha)^{n}.
\end{equation}
Integrating this relation on the interval $(0,1)$, using
(\ref{eq:border}) and the properties of the gamma function,
we obtain (\ref{eq:rel}).
\subsection{ Functions ${\tilde{S}}_{n}$}
Let us define the function
\begin{equation}\label{eq:stilda}
{\tilde{S}}_{n}(y)=
\sum_{k=0}^{\infty}\frac{(2k+n)!}{k!(k+n+1)!}y^{k+n+1}.\qquad(n\geq 0)
\end{equation}
It satisfies the recurrence relation
\begin{equation}
\partial_{y}{\tilde{S}}_{n+1}(y)=2y\partial_{y}{\tilde{S}}_{n}(y)
-(n+1){\tilde{S}}_{n}(y),
\end{equation}
which, after the change of variables $y=\frac{1-\alpha^{2}}{4}$, becomes
\begin{equation}
\partial_\alpha {\tilde{S}}_{n+1}(\alpha)=
\frac{1}{2}(1-{\alpha}^{2})\partial_\alpha {\tilde{S}}_{n}(\alpha)
+\frac{\alpha}{2}(n+1){\tilde{S}}_{n}(\alpha).
\end{equation}
It is easy to check that the expression
\begin{eqnarray}
{\tilde{S}}_{n}(\alpha)=\frac{(1-\alpha)^{n+1}}{2^{n+1}(n+1)},
\end{eqnarray}
is a solution of the above equation.
Recalling that $\alpha=\frac{\sigma}{\bar{\sigma}}$, we have
\begin{equation}\label{eq:stn}
{\tilde{S}}_{n}(\sigma,\bar{\sigma})=
\frac{1}{n+1}\frac{(\sigma-\bar{\sigma})^{n+1}}{(-2\bar{\sigma})^{n+1}}.
\end{equation}
\section{Expression for $h^{\alpha\beta}$ which turns constraints to the compact form}\label{sec:hcompact}
\cleq
Let us derive the compact expression for functions $h^{\alpha\beta}$,
defined in (\ref{eq:h1}), as
\begin{equation}
h^{\alpha\beta}(a,b)(\sigma)=
\frac{\sigma}{2}
\sum_{k=0}^{\infty}\frac{(-1)^{k}}{(k+1)!}
\int_{0}^{\sigma}d\sigma_{1}^{2}\cdots
\int_{0}^{\sigma_{k-1}}d\sigma_{k}^{2}
a^{(k)\alpha}(\sigma_{k}) b^{(k+1)\beta}(\sigma_{k}).
\end{equation}
The result is different when both variables $a$  and $b$ are $\sigma-$symmetric,
\begin{equation}\label{eq:h}
h^{\alpha\beta}(a,b)(\sigma)=\frac{1}{2}A^\alpha(\sigma)b^{\prime\beta}(\sigma),
\quad\quad A^\alpha(\sigma)\equiv\int_{0}^{\sigma}d\eta a^\alpha(\eta)
\end{equation}
and $\sigma-$antisymmetric,
\begin{equation}\label{eq:hn}
h^{\alpha\beta}(\bar{a},\bar{b})(\sigma)=\frac{1}{2}{\bar{a}}^\alpha(\sigma){\bar{b}}^{\beta}(\sigma).
\end{equation}
We will prove (\ref{eq:h}) by substituting  ${a}^{(k)\alpha} ({\sigma}_{k})$,
written as
\begin{equation}
{a^{(k)\alpha}}(\sigma_{k})=2\int_{0}^{\pi}d\eta a^\alpha(\eta)\frac{\partial^{k}}
{\partial\sigma_{k}^{k}} \delta_{S}(\eta,\sigma_{k}),
\end{equation}
into the expression for $h^{\alpha\beta}(a,b)(\sigma)$. Integrating over $\sigma_{k}$, we obtain
\begin{equation}
h^{\alpha\beta}(a,b)(\sigma)=\frac{\sigma}{2}a^{\alpha}b^{\prime\beta}
+\int_{0}^{\pi}d\eta a^\alpha(\eta)\sum_{k=1}^{\infty}\frac{2\sigma}{(k+1)!}
\partial^{k}_{\eta}[\eta b^{(k+1)\beta}(\eta)I_{k-1}(\sigma,\eta)],
\end{equation}
where $I_{k}$ is defined in (\ref{eq:ikdef}).
Note that the sum in the last expression is $\partial_{\sigma}S^{\beta}_{1}(a|\sigma,\eta)$,
where $S^{\beta}_{1}$ is defined in (\ref{eq:s1}).
Therefore, using (\ref{eq:sf}) for $x\rightarrow a$, we obtain (\ref{eq:h}).
In the case when both $\bar{a}$ and $\bar{b}$ are $\sigma-$antisymmetric,
observing that
\begin{eqnarray}
{\bar{a}}^{(k)}(\sigma_{k})=\int_{0}^{\sigma_{k}}d\eta{\bar{a}}^{(k+1)}(\eta),
&&
{\bar{b}}^{(k+1)}(\sigma_{k})=({\bar{b}}^\prime)^{(k)}(\sigma_{k}),
\end{eqnarray}
with the help of (\ref{eq:h}), we obtain
\begin{eqnarray}
h^{\alpha\beta}(\bar{a},\bar{b})(\sigma)
=h^{\beta\alpha}\Big{(}{\bar{b}}^\prime,\int\bar{a}\Big{)}=
\frac{1}{2}{\bar{a}}^\alpha{\bar{b}}^\beta.
\end{eqnarray}

\end{document}